\documentclass[10pt,pra,twocolumn,superscriptaddress,showpacs,floatfix]{revtex4}

\usepackage{graphicx}
\usepackage{dcolumn}
\usepackage{bm}
\usepackage{subfigure}
\usepackage{amssymb}
\usepackage{times}
\usepackage{amsmath}
\usepackage{amsthm}
\usepackage{color}
\usepackage[toc,page,title,titletoc,header]{appendix} 
\usepackage{framed} 
\usepackage{pdfpages}

\begin{document}

\title{Quantum hacking: saturation attack on practical continuous-variable quantum key distribution} 

\author{Hao Qin}\affiliation{Telecom ParisTech, CNRS LTCI, 46 rue Barrault, 75634 Paris Cedex 13, France}
\author{Rupesh Kumar}\affiliation{Telecom ParisTech, CNRS LTCI, 46 rue Barrault, 75634 Paris Cedex 13, France}
\author{Romain All\'eaume}\affiliation{Telecom ParisTech, CNRS LTCI, 46 rue Barrault, 75634 Paris Cedex 13, France}

\date{\today}

\begin{abstract}
We identify and study a new security loophole in continuous-variable quantum key distribution (CV-QKD) implementations, related to the imperfect linearity of the homodyne detector. By exploiting this loophole, we propose an active side-channel attack on the Gaussian-modulated coherent state CV-QKD protocol combining an intercept-resend attack with an induced saturation of the homodyne detection on the receiver side (Bob). We show that an attacker can bias the excess noise estimation by displacing the quadratures of the coherent states received by Bob. We propose a saturation model that matches experimental measurements on the homodyne detection and use this model to study the impact of the saturation attack on parameter estimation in CV-QKD.
We demonstrate that this attack can bias the excess noise estimation beyond the null key threshold for any system parameter, thus leading to a full security break. If we consider an additional criteria imposing that the channel transmission estimation should not be affected by the attack, then the saturation attack can only be launched if the attenuation on the quantum channel is sufficient, corresponding to attenuations larger than approximately 6 dB.  We moreover discuss the possible counter-measures against the saturation attack and propose a new counter-measure based on Gaussian post-selection that can be implemented by classical post-processing and may allow to distill secret key when the raw measurement data is partly saturated.
\end{abstract}

\maketitle


\section{Introduction}
\label{sec:intro}  

Quantum key distribution (QKD)~\cite{Sca09} enables two remote parties Alice and Bob to share common secure keys which are unknown to a potential eavesdropper. Unconditional security of QKD is based on the fundamental laws of quantum mechanics. Side-channels attacks nevertheless remain a crucial problem to guarantee the security of practical implementations. As a matter of fact, the models used in security proofs to describe QKD implementations may not capture all the possible deviations associated with device imperfections. This opens the possibility of attacks against QKD implementations, exploiting either passive (information leakage) or active (induced by the attacker) side-channels. 

In discrete-variable QKD (DV-QKD), various quantum hacking strategies exploiting some implementation imperfections have been proposed, and some of them demonstrated in experiments~\cite{Xu,Makarov2010,Makarov2011}. Most of the practical attacks that have been demonstrated up to now in DV-QKD consist in attacks targeting the detection part of QKD systems~\cite{Qi2007,Zhao2008,Xu,Makarov2009,Makarov2011,Weier,Wiechers2011} and exploit imperfection of single-photon detectors.

Continuous-variable QKD (CV-QKD), is another promising approach to perform quantum key distribution. It relies on continuous modulation of the light field quadratures, which can be measured with coherent detectors such as homodyne or heterodyne detections. CV-QKD  inherits several interesting features associated to the use of coherent detection instead of single-photon detectors: at the system level, CV-QKD can be implemented with off-the-shelf components that are also used and optimized in modern optical communications, allowing for a convergence between quantum and classical communications \cite{Kleis2014} and also simplifying the path and the undertaking associated to photonic integration. Coherent detectors moreover act as efficient and almost single-mode filters, leading to a superior capacity for CV-QKD to be wavelength-multiplexed with intense classical channels over WDM networks \cite{KumarWDM2015}. 

The Gaussian modulated coherent state protocol (GMCS) CV-QKD protocol \cite{GG02} is proven secure against collective attacks and recent works have shown progress in proving its security against arbitrary attacks~\cite{Anton13}. However, similarly to DV-QKD, practical CV-QKD systems can face security threats linked to imperfect implementations. The validity of security proofs indeed relies on assumptions that may be violated in practical setup, opening loopholes that may be exploited by Eve to mount attacks. For example direct~\cite{testlo} or indirect~\cite{Wave13,Wave13v2,Wave14} manipulation of local oscillator (LO) intensity can fully compromise the security. This imposes to monitor LO intensity and to use filters to forbid wavelength-dependent LO intensity manipulations. Moreover, LO intensity fluctuations also can possibly compromise the security of practical system~\cite{Loflu,Loflu2}   and  a stabilization of LO intensity is proposed to defend against such attacks~\cite{Loflu2}. 

In this work, we have identified a new loophole associated to the finite range over which coherent detectors respond  non-linearly. We have shown that it can be used to attack practical implementations of the GMCS CV-QKD protocol. Instead of targeting the shot noise calibration by manipulating the local oscillator, we propose an attack that aims at the homodyne detection located on Bob side, and more specifically at the electronics of the homodyne detection. We name our attack ''saturation attack'': it combines the induced saturation of the homodyne detection response with a full intercept-resend attack~\cite{IR}. Based on a realistic model of the homodyne detection response and saturation behavior, we can show that the saturation attack can be used get information about Alice modulated input (via intercept-resend attack, which should in theory bring the key rate to zero) while jointly manipulating the measurement results on Bob's side (taking advantage of the induced, non-linear response of the homodyne detector). For some channel and protocol parameters, the saturation attack can lead Alice and Bob to generate, at a positive rate, a key that they consider as secure, although such key will be totally insecure due to the intercept-resend attack. Hence the attack can lead to a full security break. Importantly, the attack is also practical and can be realistically launched against existing implementations, since all practical coherent detectors have a finite linearity domain and could be driven (if not monitored) outside of this domain of linearity by displacing the mean value of the received quadratures. We however propose a counter-measure that can be implemented simply, by performing on a numerical test on the measurement data. The counter-measure consists in a pre-calibration of the linearity domain of the homodyne detector and then to apply a Gaussian post-selection filter to the quadrature measurements results of Bob, so that the post-selected measurements results fall within the linearity domain while the post-selected input data is guaranteed to be Gaussian.
 
This article is organized as follows. In section \ref{pre} we first present the GMCS protocol and explain how parameter estimation is performed in this protocol. In section \ref{practical}, we shortly review existing work on the practical security of CV-QKD and propose the idea of the saturation attack in section \ref{idea}. Then in section \ref{sathomo}, we study experimentally the influence of saturation on a practical homodyne detector and propose a simple saturation model to account for it. In section~\ref{Saturation attack}, we propose a strategy to mount an active attack against the GMCS CV-QKD protocol, taking advantage of induced saturation. In section \ref{analysis}, we perform numerical simulations to analyze the influence of the saturation attack on parameter estimation, in particular on channel transmission and excess noise and then discuss the impact on secret key rate, under two different security criteria. In section \ref{Counter measure} we discuss possible counter-measure and present and analyze a counter-measure based on Gaussian post-selection. Finally, in section \ref{conclusion} we summarize the main results of our work and discuss some perspectives.
 
\section{Gaussian modulated coherent state continuous-variable quantum key distribution}\label{pre}
\subsection{Protocol}

In the GMCS CV-QKD protocol~\cite{GG02}, Alice encodes information on coherent states of light, that can be easily produced by a laser. The information is encoded on the quadratures $X_A$ and $P_A$ of coherent states; with a centered bivariate Gaussian modulation of variance $V_A \,N_{0}$. $N_{0}$ is the shot noise variance that appears in the Heisenberg uncertainty relation for the non-commuting quadratures, it corresponds to the variance of the homodyne detection output when the input signal is the vacuum field. Alice sends these Gaussian modulated coherent states, which constitute the quantum signal, to Bob through the quantum channel. On the reception side, Bob randomly chooses to measure either quadrature $X$ or $P$ by performing a balanced homodyne detection on the signal, using for that a strong phase reference, called local oscillator, and switching the quadrature measurement by varying the relative phase of LO with respect  the quantum signal to be either $0$ or $\pi/2$. 

Keeping track of modulated quadrature data $X_A$  (or $P_A$) and quadrature measurement results $X_B$ or ($P_B$), Alice and Bob obtain strings of correlated classical data by repeating many times this process over time over successive pulses. They can then use error correction to obtain identical strings from their correlated data through reverse reconciliation~\cite{GG02,Jouguet11} and further perform privacy amplification to obtain a secret key. 

In the analysis carried out in this article, that focuses on the impact of a new side-channel on CV-QKD, we don't consider finite size effects \cite{Anton10}, and assume that all the estimations are performed in the asymptotic limit. We moreover consider the security against collective attacks to compute secret key rate. One can moreover show that Gaussian attacks are the optimal collective attacks against the GMCS protocol in the asymptotic limit of infinite number of signals \cite{RN06,MFA06}. 
Hence we can analyze the security of the protocol by considering a linear channel model with additive Gaussian noise. In this Gaussian linear model, the Alice-Bob channel is fully characterized by two parameters: the channel transmission and the excess noise. The channel transmission is related to the channel loss and can be derived from the correlation between Alice and Bob's data. The  excess noise is the variance of Bob quadrature measurements in excess of the shot noise, it can be due to device imperfections (in particular imperfect modulation and noisy detections) or eavesdropper's actions on the channel.  

\subsection{Parameter estimation}
In order to estimate  parameters from Alice and Bob's correlated variables, the Gaussian linear model (Eq.(1)) with additive Gaussian noise is considered. 
 \begin{align}
  X_B=tX_A+X_N
  \end{align}
 In Eq.(1), $t=\sqrt{\eta T}$,  $T$ is the channel transmission and $\eta$ is the optical transmission through Bob set-up (including homodyne detection's finite efficiency). On Alice side, $X_A$ is a Gaussian random variable centered on zero with  variance $V_{A}$. $X_N$ is the total noise which follows a centered normal distribution with  variance $\sigma_N^2=N_0 + \eta T \xi + v_\mathrm{ele}$. This variance includes shot noise $N_0$, excess noise $\xi$ and electronic noise of Bob $v_\mathrm{ele}$. 
 
 
 In this article, we follow the parameter estimation procedure of Ref~\cite{Jouguet12a}.
 We can obtain  three equations (Eq. (2)-(4)) relating modulated data $X_A$ and measured data $X_B$ to parameter estimation :
  \begin{align}
 V_A &=Var(X_A)= \langle (X_A-\langle X_A \rangle)^2\rangle \\
 V_B &=Var(X_B)= \langle (X_B-\langle X_B \rangle)^2\rangle\\ \nonumber
 &=\eta T V_A + N_0 + \eta T \xi + v_\mathrm{ele}\\
  Cov(X_A,X_B)  &=\langle X_AX_B\rangle-\langle X_A\rangle\langle X_B\rangle \\  \nonumber
    &= \sqrt{ \eta T} V_A
   \end{align}

Additionally, in order to measure the shot noise $N_0$,  Bob needs to close the signal port so he can measure the variance when the input signal is vacuum. When there is no signal impinging on the homodyne detection, the variance of homodyne detection is used to calibrate the value of shot noise. In this case Eq. (3) reduces to an additional equation, obtained by performing a shot noise calibration:
 \begin{align}
   V_{B_{0}} = N_0 + v_\mathrm{ele}.
 \end{align}
Note that $\eta$ and $v_\mathrm{ele}$ are also calibrated values, measured before launching the protocol. \\

The parameter characterizing the quantum channel in the Gaussian linear model, i.e. $T$ and $ \xi$ can then be estimated from Eq. (2)-(4):

\begin{equation}
 T =\dfrac{Cov(X_A,X_B)^{2}}{\eta Var(X_A)^{2}}
\end{equation}
 \begin{align}
 &\xi =\dfrac{Var(X_B)}{\eta T}-Var(X_A)-\dfrac{N_0}{\eta T}-\dfrac{v_\mathrm{ele}}{\eta T}
 \label{eqexn}
\end{align}
 Additionally, by calibrating the shot noise variance  $N_0$ from Eq.(5), all variances and correlations can be normalized in shot noise units and can then be used to estimate the secret key rate.

\subsection{Security model and achievable secret key rate}
In order to estimate the secret key rate, Alice and Bob need to compute the mutual information between their data and estimate an upper bound of Eve's information. In this article, parameter estimation and secret key rates will be analyzed in the context of collective attacks, in the asymptotic regime \cite{RN06}. Although the security of CV-QKD can be analyzed in a more general setting, we want to stress that extending our analysis to more general (and complex) security models would not qualitatively change the main finding of our article. As a matter of fact we exhibit an explicit attack strategy, exploiting the saturation of the homodyne. As we shall demonstrate, this attack leads to a complete security break against a an attacker limited to collective attacks, assuming parameter estimation is performed in the asymptotic regime. By extension, our proposed attack would also lead to a complete security break under more general security models, that consist in ``increasing'' the power of the eavesdropper.

A lower bound on the secret key rate achievable against collective attack (in the asymptotic limit) for the CV-QKD protocol can be expressed as $R=\beta I_{AB}-\chi_{BE}$ \cite{Lodewyck07}. It is composed of two terms:  $I_{AB}$ is the mutual information between Alice and Bob and $\chi_{BE}$  the Holevo bound of Eve's knowledge,  $\beta\in [0,1]$  is the reconciliation efficiency, related to the fact that practical error correction usually does not reach the Shannon limit (which would correspond to the case $\beta = 1$). $I_{AB}$ is a decreasing function of the excess noise, while $\chi_{BE}$ is an increasing function of excess noise, hence any rise of the excess noise will lead to a decrease of the secret key rate $R$.

\section{Practical security issues: loopholes and attacks in CV-QKD}
\label{practical}

In practical CV-QKD implementations \cite{Lodewyck07, Jouguet13}, the local oscillator  is transmitted publicly on the optical line between Alice and Bob, multiplexed with the quantum channel. Hence the LO can be accessed, and thus manipulated by an attacker in practical implementations. It is important to note that  the LO can in principle be generated locally at Bob side, as demonstrated in recent proof-of-principle experiments \cite{BQi2015, Soh2015}, where the LO is phase-locked with the quantum signals emitted by Alice. However, phase-locking two distant lasers brings more complexity and noise and all practical CV-QKD full demonstrations have so far been performed with a "public" LO. This opens the door to different attack strategies  based on LO manipulation. An eavesdropper can for example modify several properties of  the LO pulse,  such as the intensity, the wavelength, or the pulse shape  \cite{testlo,Jouguet13lo,Wave13,Wave13v2,Loflu,Loflu2,Wave14}. The eavesdropper can in particular bias the shot noise calibration (Eq. (5)) by manipulating the LO intensity or its overlap with the quantum signal.  We have indeed seen that the excess noise is expressed in shot noise units. If the shot noise is overestimated while all the other measurements remain unchanged, the excess noise in shot noise unit (SNU) will then be underestimated. As a consequence Alice and Bob will then overestimate their secret key rate, leading to a security problem. 

  Most existing attacks rely on shot noise estimation induced by different LO manipulations combined with specific strategies. In Ref~\cite{testlo}, equal-amplitude attack is described. By replacing the quantum signal and the LO pulse by two squeezed states of equal amplitude, Eve can make Alice and Bob measure an excess noise estimate that is much lower than the actual shot noise. This attack may allow Eve to break the security without being detected if Bob doesn't monitor the LO intensity, that is strongly modified in this attack. In Ref~\cite{Jouguet13lo}, the authors propose a strategy where Eve changes the shape of the LO pulse to introduce a delay on the clock trigger. As a consequence, the variance of the shot noise measurement can be lowered without changing the LO power. Such calibration attack biases the estimation of shot noise and thus the excess noise in shot noise units. The authors propose a counter-measure based on real time monitoring of the shot noise method to prevent this LO manipulation loophole. In Ref~\cite{Wave13} and Ref~\cite{Wave13v2}, as an extension of the equal-amplitude attack~\cite{testlo}, a wavelength attack on a CV-QKD system using heterodyne detection has been proposed. In this attack, by exploiting the wavelength dependent property of the homodyne detection's beam splitter, Eve can bias the intensity transmissions of LO and signal. By inserting light pulses at different wavelengths, this attack allows Eve to bias the shot noise estimation even if it is performed in real time. This attack can be prevented by adding a wavelength filter before the beam splitter. Recently  in~\cite{Wave14}, similar wavelength attack  has been proposed to compromise the practical security of CV-QKD system using an homodyne detection. An improved real-time shot noise measurement technique is also proposed to detect this attack, closing all known wavelength attack loopholes.
 
 To summarize,  the main idea of existing attacks on CV-QKD consist in manipulating the local oscillator in different ways so that the eavesdropper can bias the shot noise estimation and thus the excess noise. The threat of such attacks can be removed if Alice and Bob "locally" generate LO pulse \cite{BQi2015, Soh2015}  or measure the shot noise in real time instead of relying on an off-line calibration \cite{Jouguet13lo}. LO is an important issue for the practical security in CV-QKD, but as we will demonstrate with the introduction of the saturation, it is not the only implementation loophole that should be considered in practical CV-QKD.

 \section{ Principle of the saturation attack }
 \label{idea}

 Unlike the attacks aiming at the local oscillator, we introduce a new attack on CV-QKD which exploits the finite linearity domain of the homodyne detection response. Indeed, an implicit but nevertheless fundamental assumption in the security proofs of CV-QKD is that the response of the homodyne detection is linear with respect to the input quadrature. This assumption is necessary because parameter estimation (Eq. 2-4) assumes that the quadrature measurement performed by Bob are linearly related to the optical field quadratures, in order to relate them to the parameters ($T$ and $\xi$) of the quantum channel.  However, for a practical coherent detector, such as the homodyne detection used in to implement the GMCS CV-QKD protocol, the linearity domain is limited. If the value of input quadrature is too large, linearity may not be verified, leading to a saturated behavior.

From section II.B, we  can observe that, based on the Gaussian linear model (Eq.1), the parameter estimation consists in the evaluation of the  covariance matrix. It is interesting to notice that the different coefficient of the covariance matrix are invariant under any linear shifts of the quadratures. Indeed the security evaluation in CV-QKD relies solely on the evaluation of second order moments of the quadrature, while the first order moments (mean value) are not monitored. This leaves Eve the freedom to manipulate quadratures mean value. Combining this observation with the existence of a finite domain of linearity for the detection, a natural attack strategy for Eve is to actively introduce a large displacement on the quadrature received by Bob in order  to force the  homodyne detection to operate in its saturated region. This strategy, that is the core idea of the saturation attack, enables Eve to influence Bob's measurement results and to bias parameter estimation. Importantly, unlike  the attacks targeting the local oscillator, in which the shot noise measurement is influenced, saturation attacks do not bias the shot noise calibration but still influence the excess noise estimation.

 \section{Saturation of a homodyne detector}\label{sathomo}

\begin{figure}[t]
    
           \centering
     {\includegraphics[width=0.4\textwidth]{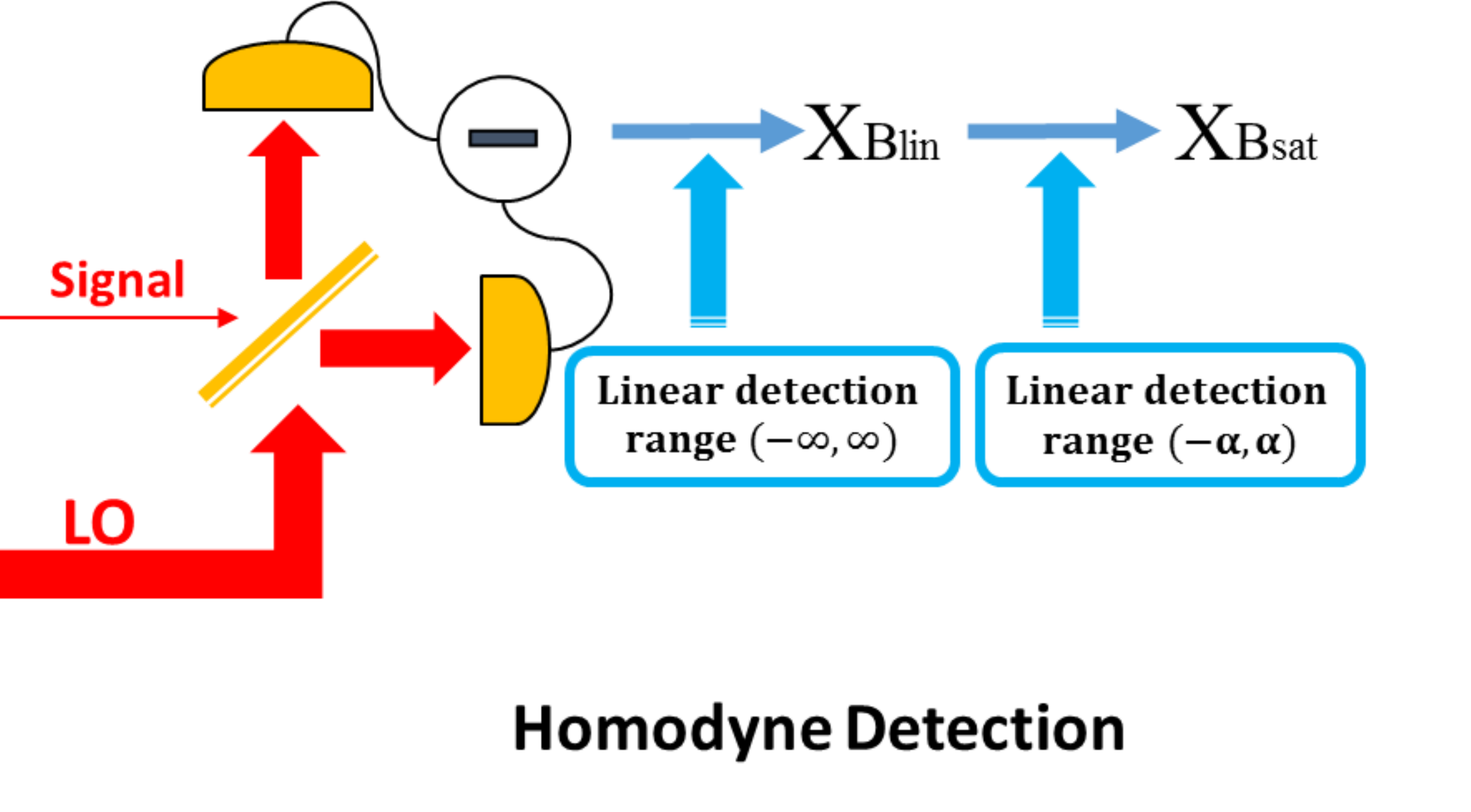}}
           \caption{Model for a practical homodyne detection:  its output $X_{B_{sat}}$ can be seen as the ideal output $X_{B_{lin}}$ on which a saturation function is applied (Eq. \ref{eqsatd})}
           \label{homodyne}
        \end{figure}

Saturation of the homodyne detection typically occurs when the input field quadrature overpasses a threshold. This threshold depends on characteristics of the homodyne detector's electronics, such as the amplifiers linearity domain or the data acquisition card (DAQ) range (Fig.\ref{homodyne}). If Bob performs quadrature measurements on input  signals  falling outside of the detector's  linearity range, the measurement statistics will be influenced by the saturation. Saturation will in particular lead to decrease in the variance of the measurement results.   
   
  \subsection{Saturation model}\label{model}
  
 The quadrature measurement  performed with an homodyne detection consists in the subtraction in the electronic domain of the photo-currents produced by the two photodiodes followed by an electronic front-end and acquisition. The standard analysis considers that the homodyne response is linear with respect to the input quadratures. We then denote the measured quadrature as $X_{B_{lin}}$ ($X_B$ in section.II).  However, the linear detection range of a practical homodyne detector  cannot be infinite. We propose a saturation model (Eq.\eqref{eqsatd}) with predefined upper and lower bounds for the homodyne detection response: for quadrature input values between these two bounds, the response of homodyne detection is unaffected, otherwise it saturates to a constant value.  To simplify the analysis, we have assumed in this model that the linear detection range can be described by one single parameter, $\alpha$, intrinsic to the detector.  Under this saturation model, the linear range is $[-\alpha,\alpha]$ and the measured quadrature is called $X_{B_{sat}}$. The relation between $X_{B_{sat}}$and $X_{B_{lin}}$ is the following: 
  \begin{equation}
  \begin{split}
       &\quad\quad X_{B_{lin}} \geqslant \alpha,\quad \quad \quad \quad X_{B_{sat}} =\alpha        \\
   if  &\quad\mid  X_{B_{lin}} \mid < \alpha, \quad then  \quad  X_{B_{sat}} = X_{B_{lin}}  \\
       &\quad\quad X_{B_{lin}} \leqslant -\alpha,\quad \quad \quad  X_{B_{sat}} =-\alpha \qquad \qquad  \     
   \end{split}
   \label{eqsatd}
   \end{equation}
 As expected, if $\alpha\rightarrow \infty$, the saturation model is equivalent to the standard linear model. In a typical (non saturated) CV-QKD implementation, the value of $\alpha$ is large enough to ensure that field quadratures almost never overpass the saturation threshold limit. Alice and Bob can in practice guarantee the linearity by limiting number of photons impinging on the homodyne detector to be much smaller than $\alpha^2$.  Since the limit $\alpha$ is intrinsic to the electronics of the detector, a practical way to guarantee with high probability that  $\alpha \gg X_{B_{lin}}$ is to lower the LO intensity so that the shot noise value $N_0 \ll\alpha^2$. In general, input quadrature modulation variance are calibrated in shot noise units which depends on LO intensity and Alice can choose  a Gaussian modulation with $\langle X_{B_{lin}}\rangle=0$ and $Var(X_{B_{lin}}) \ll \alpha^2$ so that the detector does not saturate. However, as mentioned earlier this procedure cannot cope with situations where $X_{B_{lin}}$ mean value is strongly displaced, as it will be the case in saturation attack.

\subsection{Experimental observation of saturation}
In a practical  balanced homodyne detector, the common mode rejection ration (CMRR) is not infinite and the mean value of the homodyne detection in absence of input signal is affected by the imbalance, leading to: $\langle X_{B_{0,lin}}\rangle=\epsilon  I_{LO}$, where  $I_{LO}$ is the LO intensity, and $\epsilon$ is the imbalance factor which is dependent on experimental imperfections such as  photodiode quantum efficiency mismatch or  beam-splitter imbalance.

Because of this imperfection (but in absence of saturation), the relation between measured noise variance (in $Volt^2$) and LO intensity (in $ \mu W$) usually can be written as: $Var(X_{B_{0,lin}})=AI_{LO}+B$~\cite{Chi11} (We neglect the quadratic part since in our case the  LO power is relatively low). $I_{LO}$ is the LO intensity,  $A$ is linear with $I_{LO}$ and is  related to shot noise while  $B$ is independent of $I_{LO}$ and is related to electronic noise. The value of  $A$ and $B$ can be determined experimentally. 

\begin{figure}
\centering
\subfigure[Mean of the homodyne output $\langle X_{B_{0,sat}}\rangle $vs LO Intensity.]{
\begin{minipage}[b]{0.45\textwidth}
\includegraphics[width=1\textwidth]{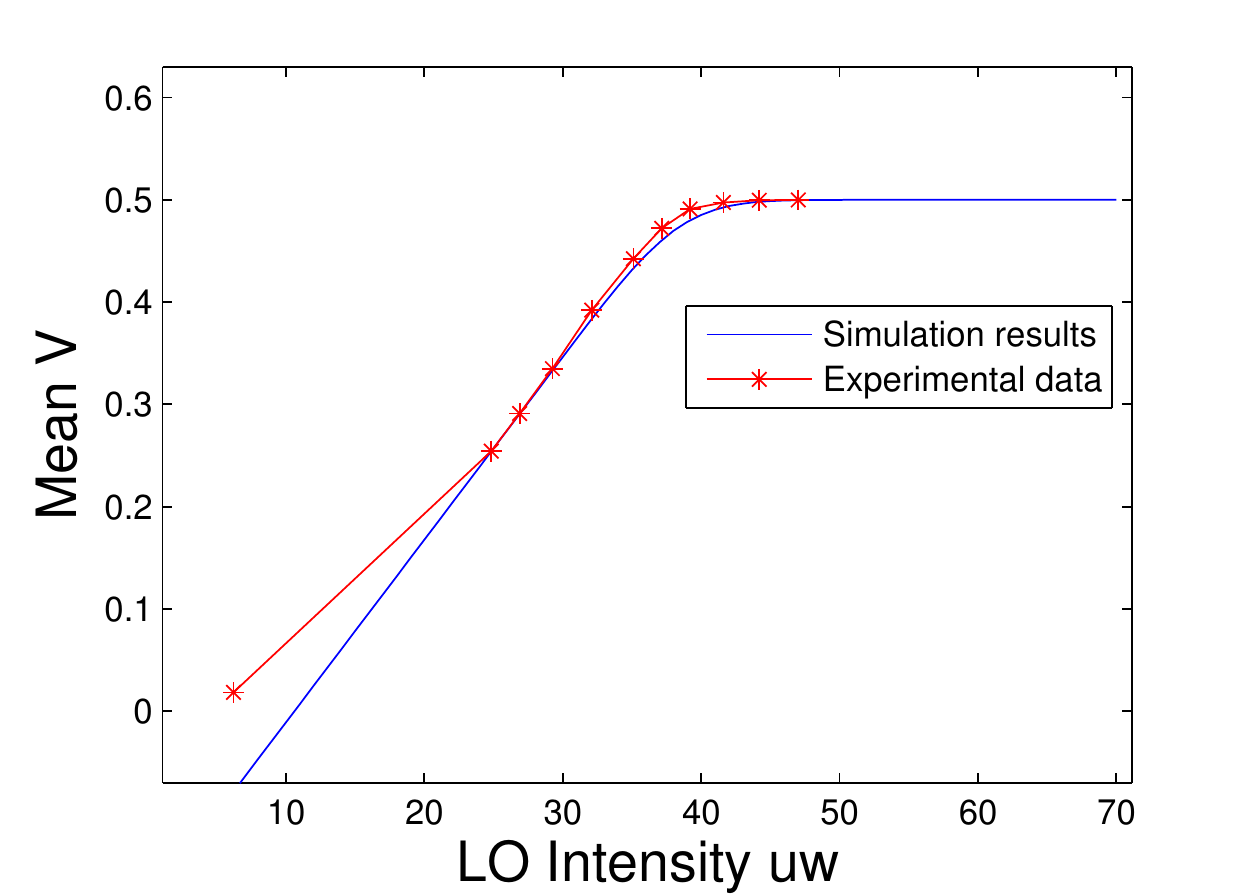}
\end{minipage}
}
\subfigure[Variance of the homodyne output $Var(X_{B_{0,sat}})$ vs LO Intensity.]{
\begin{minipage}[b]{0.45\textwidth}
\includegraphics[width=1\textwidth]{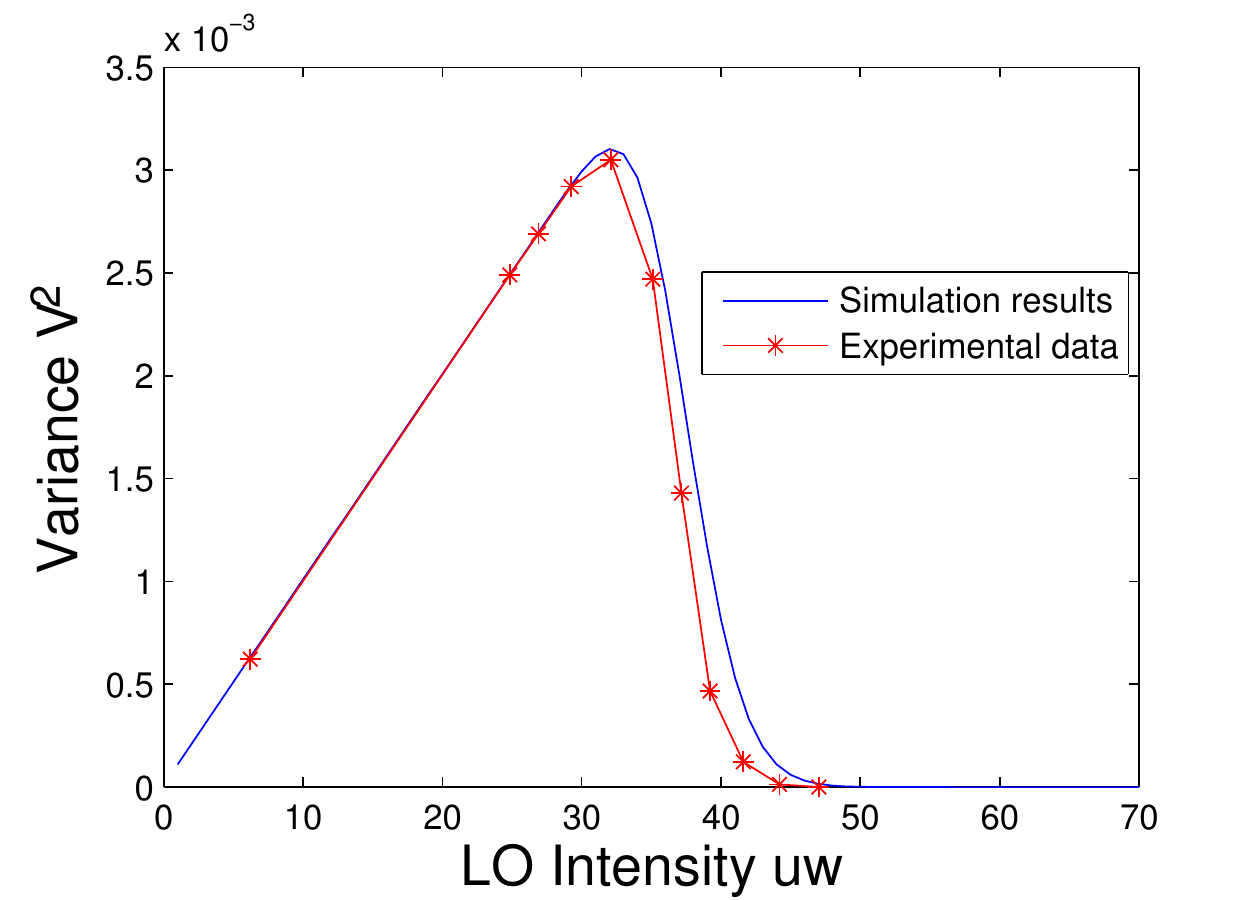}
\end{minipage}
}
 \caption{ Experimental characterization of the saturation behavior of a practical homodyne detection.} \label{mean}
\end{figure}

We have performed experimental shot noise measurement, measuring the variance of the homodyne detection output, as a function of $I_{LO}$. This has revealed that the  measured shot noise variance is linear with $I_{LO}$ on a given range, and then drops when the LO intensity is above a certain value. We have analyzed this behavior with the saturation model presented in section V.A and compared its prediction to experimental measurements of Fig.\ref{mean}.  We display the measured  homodyne detection output and its variance,  for vacuum input signal as a function of $I_{LO}$.  The experimental results are displayed on Fig.\ref{mean}. The linear behavior can be observed when LO intensity is below 35 $\mu W$. Due to the imbalance of homodyne detection ($\epsilon$), the mean value of the homodyne output can become large as the LO intensity increases. If these values overpass the linearity threshold  (in the present case 0.5 V, due to the DAQ card) the homodyne detection response saturates to a constant value (Fig.\ref{mean} (a)). As a consequence, measured shot noise variance strongly decreases (Fig.\ref{mean} (b)) when such saturation happens. 

In order to check the validity of the saturation model introduced in Eq.\eqref{eqsatd}, we have simulated the expected homodyne detection response with our saturation model and compared it with experimental measurements. We first determine the parameters $A$ and  $B$ from linear fit on $\langle X_{B_{0,lin}}\rangle$ and $Var(X_{B_{0,lin}})$ versus LO intensity over the domain of linearity ($I_{LO}<35\mu W$). The saturation parameter $\alpha$ is here fixed by our DAQ range: $\alpha=0.5 V$. We then apply the saturation model Eq.\eqref{eqsatd} to the  variable $X_{B_{0,lin}}$ to obtain  $X_{B_{0,sat}}$. We compute the  mean $\langle X_{B_{0,sat}}\rangle$ and the variance $Var(X_{B_{0,sat}}^2)$, which result in the behavior  shown in Fig.\ref{mean}. For the measured shot noise under saturation, the simulation results match very well with our experimental data. It indicates that our proposed saturation model is realistic and can be further used to interpret our saturation attack.

\section{Attack strategy}\label{Saturation attack}

 \subsection{Intercept-resend attack}
The intercept-resend attack play an important role as being one part of our saturation attack. Intercept-resend attack~\cite{IR} in CV-QKD is achievable with today's technologies and its security analysis has been studied in previous work~\cite{IR}.  In such attack, Eve intercepts all the pulses sent by Alice on the quantum channel and measures simultaneously the $X$ and $P$ quadratures with the help of a heterodyne detection. Eve then prepares a coherent state according to her measurement results and sends it to Bob. Under such attack the correlation between Eve and  Bob data will be stronger than the one between Alice and Bob so that Eve always has an information advantage. Due to measurement disturbance and coherent state shot noise, the  intercept-resend attack, that is entanglement-breaking,  introduces two shot noise units of excess noise. In practice, Eve's device and her action can introduce additional excess noise. A full intercept-resend attack will therefore introduce at least two shot noise units of excess noise, which should forbid the generation of secret key under collective attacks. This however assumes that the estimation procedure is not biased. We will see on the  contrary that a saturation attack can bias the excess noise estimation and lead to a security break.

\subsection{Saturation attack}
\label{saturationattack}
\begin{figure*}[t]
 
        \centering
     {\includegraphics[width=0.8\textwidth]{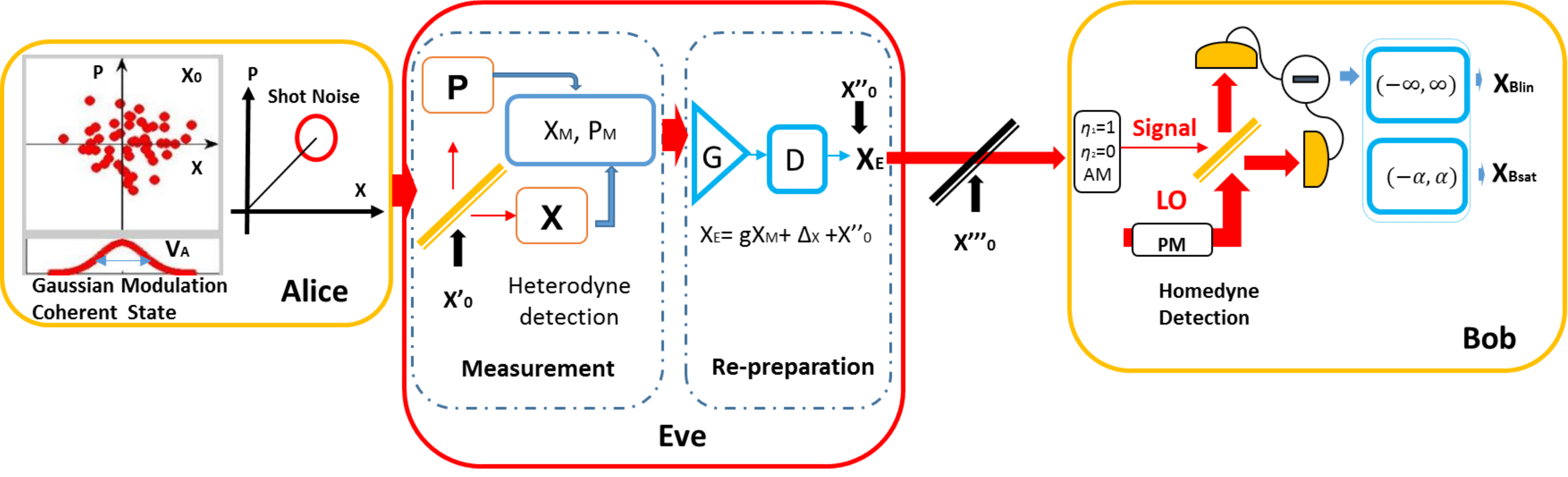}}
      
        \caption{General description of GMCS CV-QKD under the saturation attack. Alice: prepare the coherent state with quadratures X and P; Eve: measurement and re-preparation stage, G:gain, D:displacement; Bob: perform the homodyne detection, AM:amplitude modulator, $\eta_1,\eta_2$: signal transmission coefficients, PM:phase modulator, $-\alpha, \alpha$: linear working range.}
        \label{Eve}
     \end{figure*}

The saturation attack on the GMCS CV-QKD protocol is an active attack, where Eve combines a full intercept-resend attack with an induced saturation of Bob detector. Saturation of Bob homodyne detection is obtained by displacing the quadrature of the resent coherent state. The displacement value $\Delta$ is chosen by Eve but is constant for each resent coherent state pulse. When performing the intercept-resend attack, Eve can also choose to rescale the resent quadratures by a gain $g$.

Eve chooses the attack parameter  bias the estimated excess noise  below the null key threshold (calculated under collective attack~\cite{Lodewyck07}), so that according to their estimation, Alice and Bob will assume they can obtain a positive key rate, and will accept to distill such a key based on parameter estimation, while no secure key can be obtained from the actual correlations since a full intercept-resend attack has been performed. 

We propose a visual description of our saturation attack in Fig.\ref{Eve} in which we distinguish mainly two functional blocks: quadrature measurement and quadrature re-preparation. 
 By using a heterodyne detection, Eve measures Alice's quadrature $X_A$ and $P_A$ simultaneously. 
 In order to simplify our analysis, we assume that Eve's station is located at  Alice's output  and that the channel transmission between Alice-Bob and Eve-Bob are equal. Moreover, we assume that Alice and Bob measure their shot noise and monitor the LO intensity  in real time~\cite{Jouguet13lo}, with two transmission coefficients randomly decided at Bob side ($\eta_1=1,\eta_2=0$).

Eve's measurement results ($X_M, P_M$) after the heterodyne measurement are expressed as: 
 \begin{align}
X_M=\dfrac{1}{\sqrt 2}(X_A+X_0+X_0'+X_{N_{A,E}})\\
P_M=\dfrac{1}{\sqrt 2}(P_A+P_0+P_0'+P_{N_{A,E}})
\end{align}

Where $X_0$ is a noise term due to the  coherent state  encoding of Alice  while $X_0'$ is a noise term due to 3 $dB$ loss in the heterodyne detection. $X_{N_{A,E}}$ is a random noise that accounts for the technical  noise of Alice's preparation and Eve's measurement process with its variance $\xi_{A,E}$.

In the re-preparation stage, Eve  prepares a coherent state with quadratures ($X_E, P_E$) according to her measurement ($X_M, P_M$). Eve can also induce a displacement ($\Delta_X, \Delta_P$) and an  amplification ($g$) on the data $X_M$ before optical encoding.  In  our further analysis, we only look at the $X$ quadrature but the treatment for the quadrature $P$  is  totally symmetric.  The resend quadrature of  Eve can be written as:  

\begin{align}
  X_E&=gX_M+\Delta_X+X_0''\\
  &=\dfrac{g}{\sqrt 2}(X_A+X_0+X_0'+X_{N_{A,E}})+\Delta_X+X_0''
  \end{align}
  
Where, $X_0''$ is a noise term  due to coherent state encoding of Eve.   $X_0$, $X_0'$ and $X_0''$ all follow $\mathcal N (0,N_0)$ with their variance equal to one unit of shot noise ($N_0$). 
  
Introducing displacement on coherent states is experimentally achievable~\cite{dis}  and since Eve prepares the new states, the displacement parameter  ($\Delta_X, \Delta_P$) can be freely chosen by her. We will first consider that Eve chooses an amplification coefficient $g=\sqrt 2$, in order to compensate the loss from the heterodyne detection.  \\

\paragraph{Linear detection}

On Bob side, Bob measures the quadrature sent by Eve by performing a homodyne detection. We first consider Bob uses a homodyne detection whose linear detection range is infinite (Fig.1). The measured quadrature ($X_{B_{lin}}$) can be written as   
\begin{equation}
X_{B_{lin}}=t(X_E+X_{N_{E,B}})+\sqrt{1-t^2}X_0'''+X_{ele}
\label{xblin}
\end{equation}

After the propagation through the lossy channel,  technical noise of  Eve and Bob $X_{N_{E,B}}$ ($Var(X_{N_{E,B}})=\xi_{E,B}$), vacuum noise $\sqrt{1-t^2}X_0'''$ ($Var(X_0''')=N_0$) and  electronic noise of Bob $X_{ele}$ ($Var(X_{ele})=v_\mathrm{ele}$) are added to the quadrature prepared by Eve ($X_E$). Here  $t=\sqrt{\eta T}$, where $T$ is the channel transmission between Eve and Bob, and $\eta$ is the Bob's efficiency. The correlation between Alice and Bob and the variance of Bob can be written as:

\begin{align}
\begin{split}
 Cov(X_A,X_{B_{lin}})&=\langle X_AX_{B_{lin}}\rangle\\
 &=\dfrac{tg}{\sqrt 2}\langle X_AX_{A}\rangle+t\Delta_X\langle X_A\rangle\\
 &=\dfrac{tg}{\sqrt 2}Var(X_A)
 \label{eqcn}
 \end{split}
\end{align}

\begin{align}
\begin{split}
 Var(X_{B_{lin}})=& \langle X_{B_{lin}}^2\rangle -\langle X_{B_{lin}}\rangle ^2\\
= &\dfrac{t^2g^2}{2} [Var(X_A)+2N_0+\xi_{sys}]+(1-t^2)N_0\\
&+t^2N_0+v_\mathrm{ele}+t^2\Delta_X^2-t^2\Delta_X^2\\
= &\eta T\dfrac{ G}{2} Var(X_A)+ \eta T\dfrac{G}{2}(2N_0+\xi_{sys})\\
&+N_0+v_\mathrm{ele}
\end{split}
 \label{eqvbn}
\end{align}

In Eq. \eqref{eqcn},\eqref{eqvbn}, we can see that with an ideal linear detection range, the induced displacement $\Delta_x$ has no influence on the measurement results, since the terms of $\Delta_x$ has no impact on both correlation and variance measurements.

Under linear detection and intercept-resend attack with the gain $G=g^2=2$, the correlation (Eq.\eqref{eqcn}) is not modified by Eve's action, so that the estimated channel transmission is not biased ($T_{lin}=T$). Based on  Eq.(7), the  excess noise estimation on  Alice side is $\xi_{lin}=2N_0+\xi_{sys}$,  where $\xi_{sys}=\xi_{A,E}+\frac{2}{G}\xi_{B,E}$. Similarly to section.\ref{pre}, we introduce the noise variable $X_N$ which contains all the noise added to Bob's measurement, the variance of $X_N$ is $\sigma_N^2=\eta T\frac{G}{2}(2N_0+\xi_{sys})+N_0+v_\mathrm{ele}$. \\

 \paragraph{Saturated detection}

As we have seen the linearity of the homodyne detection cannot be guaranteed over an arbitrary large detection range. A more realistic model consists in taking saturation into account, according to the saturation model described in Eq.\eqref{eqsatd}: we denote  $X_{B_{sat}}$ the quadrature measured by Bob in a model taking saturation into account. Under this modified model,  the quadrature measured by Bob.  $X_{B_{sat}}=X_{B_{lin}}$  only if  $|X_{B_{lin}}| < \alpha $. Otherwise the quadrature measurement results saturate to a constant value, equal to the detection limit $\alpha$ or $-\alpha$.

 Eve can freely set the displacement value ($\Delta_X, \Delta_P$) so that $X_{B_{lin}}$ can partially overpass the linear range $[-\alpha,\alpha]$. In further analysis, we  consider $\Delta=t\Delta_X$  as the displacement value. In order to induce a given value of $\Delta$ on the quadrature of the coherent state impinging on Bob, Eve can choose a proper $\Delta_X$ once she knows $t$, that typically depends on the distance. As we shall see, parameter estimation affected by saturation can lead to excess noise below the null key threshold.  
 In the next section, we will show that under certain  conditions of our attack strategy, Eve can manipulate the channel transmission and the excess noise estimated by Alice and Bob, so that her intercept resend action can remain under cover while fully compromising the practical security of the CV-QKD protocol.

\section {Security Analysis}
\label{analysis}

\subsection{Parameter estimation under the saturation attack}

The channel transmission and  excess noise estimation fully characterize the quantum channel of CV-QKD, we thus only need to analyze the impact of saturation on these two estimated parameters.  It is  in particular critical to evaluate whether an induced saturation can reduce the excess noise estimation as thus opens the door to severe attacks. 

\subsubsection {Channel transmission estimation}
Under the saturation attack, Alice encodes $X_{A}$ and Bob measures $ X_{B_{sat}}$ (Eq.\eqref{eqsatd} and Eq.\eqref{xblin}) and they evaluate the correlation coefficient: $Cov(X_A,X_{B_{sat}})$ (calculation details can be found in Appendix A). From this correlation coefficient(A.3), the estimation of the channel transmission under saturation attack $\hat T_{sat}$  can be expressed as: 

 \begin{align}
 \hat T_{sat} =T\dfrac{G}{8} [1+\mathrm{erf}(\dfrac{\alpha-\Delta}{\sqrt{2  Var(X_{B_{lin}})}})]^2 (\Delta>0)
 \label{eqt}
\end{align}

In which, $\mathrm{erf}$ is the error function defined in Eq.\eqref{eq:covsat} and $Var(X_{B_{lin}})$ is  the variance of Bob's measurement under linear detection. As we have discussed in section III, a reasonable assumption for the detector linearity limit $\alpha$  is that $\alpha^2>>Var(X_{B_{lin}})$ and $\alpha^2>>N_0$, so that the measurement results of Bob  would not affected by saturation in absence of displacement.  This agrees with Eq.\eqref{eqt}: if  $\alpha-\Delta$ is much larger than $\sqrt{2 Var(X_{B_{lin}})}$, then $\hat T_{sat} \simeq T\dfrac{G}{2}$ which is the estimated value under the linear model ($G=2$ being the natural rescaling choice to compensate the loss introduced by the heterodyne detection). However when $\Delta$ is close to  $\alpha$, the impact of saturation  becomes important, and $\hat T_{sat}$ becomes smaller than $T$. An extreme case is  when  $\Delta$ is much larger than $\alpha$,  the error function  becomes $-1$ and  $\hat T_{sat}=0$.

 \subsubsection {Excess noise estimation}
From Eq.\eqref{eqexn},  the estimated excess noise  depends on the variance of Bob's measurement and on the channel transmission between Alice and Bob. Under the  saturation attack, these two values will both decrease.  We need to evaluate these two values to see whether the induced saturation will result in reducing the estimated excess noise.  We have already analyzed $\hat T_{sat}$ in the previous subsection (Eq.\eqref{eqt}). With Eq.\eqref{eqsatd}, we can  calculate  $ Var(X_{B_{sat}})$ under saturation attack (calculation details can be found in Appendix.B.). Based on $\hat T_{sat}$ and $ Var(X_{B_{sat}})$, we are able to express the estimated excess noise in shot noise units under the saturation attack :
   \begin{equation}
      \begin{split}
\dfrac{\hat \xi_{sat}}{N_0}= &\dfrac{1}{\eta T\frac{G}{2}(1+A)^2N_0}[  Var(X_{B_{lin}})(1+A-\dfrac{B ^2}{\pi})\\&-2\sqrt{\frac{2 Var(X_{B_{lin}})}{ \pi}}(\alpha-\Delta)A*B\\
&+(\alpha-\Delta)^2(1-A^2)-4N_0-4v_\mathrm{ele}]-\dfrac{V_A}{N_0}
\end{split}
  \label{eqex}
  \end{equation}
  In which $A$ and $B$ are given by:
   \begin{equation}
      \begin{split}
A = & \mathrm{erf}(\dfrac{\alpha-\Delta}{\sqrt{2 Var(X_{B_{lin}})}}),\quad   B=e^{-\frac{(\alpha-\Delta)^2}{2 Var(X_{B_{lin}})}}.
  \end{split}
\end{equation}

From Eq.\eqref{eqex}, we can verify that when the value of $\alpha-\Delta$ is much larger than $\sqrt {2 Var(X_{B_{lin}})}$,  then $A\rightarrow 1$ and  $B\rightarrow 0$, so that $\hat \xi_{sat}=\frac{Var(X_B)}{\eta  T}-Var(X_A)-\frac{N_0}{\eta  T}-\frac{v_\mathrm{ele}}{\eta  T}= \xi_{lin}$(Eq.\eqref{eqexn}). It can  be considered that  no saturation is induced and  the excess noise estimation is not affected.

\subsubsection{Estimated excess noise can be made arbitrary small}

\label{proof}

We can  prove,  by the use of the intermediate value theorem,  that  $\hat \xi_{sat}$ can be manipulated to be any value below $\xi$ and in particular any value below $\xi$ an arbitrarily small.\\

\paragraph*{Proposition}
\label{proposition}
Under saturation attack,  for any $0<\xi_T<\xi$, there always exist a particular value of the displacement $\Delta_T$  for which $\hat \xi_{sat}=\xi_T$ \\

\paragraph*{Proof}

$\hat \xi_{sat}$ is a function of $\Delta$. When $\Delta=0$,  $\hat \xi_{sat} (0) =\xi > 0$, where $\xi=2N_0+\xi_{sys}$ under intercept-resend attack. When $\Delta=2\alpha$, since we can assume that $\alpha^2>>Var(X_{B_{lin}})$, we then have $A =-\mathrm{erf}(\frac{\alpha}{\sqrt{2 Var(X_{B_{lin}})}})=-1$, and $\hat \xi_{sat} (2\alpha) \rightarrow -\infty$. Since  $\hat \xi_{sat}$  is a continuous function of $\Delta$ over the interval $[0, 2\alpha)$, then for any $\xi_T$ in  ($-\infty$,$\xi$] there always exists a $\Delta \in [0, 2\alpha)$ so that  $\hat \xi_{sat}=\xi_T$.

 \subsection {Defining criteria of success for the saturation attack}
 
 \label{definecriteria}
 
 Alice and Bob  estimate the key rate  based on their estimation of excess noise and channel transmission. If the excess noise is too large, it won't allow Alice and Bob to distill any secret key.  
 A full security break consists in an attack where Eve has full knowledge on the generated key while Alice and Bob still accept this compromised key  material.  
 An intercept-resend attack is an attack strategy that leads in general to a denial of service but not to a full security break on CV-QKD.  On the other hand, we want to claim that the saturation attack can be used to obtain a  full security break. 
 
 To clarify what we mean, we define a first criteria  (level I) for a successful saturate attack, corresponding to a set of conditions to meet: 
 
\begin{framed}
Level I criteria for a successful saturation attack:
 \begin{itemize}
 \item The attacker Eve performs the saturation attack: Intercept-resend attack combined with displacement.
  \item Alice and Bob obtain a positive key rate from their estimated parameters $\hat T_{sat},\hat \xi_{sat}$. 
 \end{itemize}
\end{framed}

This set of condition corresponds to a full security break because Alice and Bob will obtain a positive key rate under the attack and thus accept key material, while this key is insecure as it can be fully obtained by Eve.
Because the proposition enounced in \ref{proof}, we can prove the saturation attack can always meet  Level I criteria: for any quantum channel, characterized by $T$ and $\xi$, the saturation attack can turn the parameter estimation always turn the estimated parameter to $\hat \xi_{sat} \simeq 0$  while $0< \hat T_{sat}<T$. In particular, under saturation, as the estimated excess noise can be made arbitrarily close to zero, Alice and Bob will always generate some positive key rate, and level 1 criteria can always be met.

While level I criteria defines conditions for a successful attack, the induced saturation can in practice strongly decrease the estimated  channel transmission  $\hat T_{sat}$ (Eq.\eqref{eqt}). This  can be a problem in practice since Alice and Bob usually have a good \textit{a-priori} estimate of the channel transmission based on their knowledge of the channel length and of the fiber attenuation coefficient. In addition, channel loss are usually calibrated before any new optical device (such as a QKD system) is installed. If the measured channel transmission is much lower than the expected value for a given link distance, Alice and Bob can reasonably be suspicious and they may decide to reject the key. This motivates us to introduce additional conditions to the list, and to define an Level II criteria for a successful saturation attack.


\begin{framed}
Level II criteria for a successful saturation attack:
 \begin{itemize}
 \item The attacker Eve performs the saturation attack (Intercept-resend attack on each pulse combined with  displacement of each resent pulse).
 \item  Maintain the channel transmission estimation unaffected ($\hat T_{sat}=T$).
 \item Alice and Bob  obtain a positive key rate from their estimated parameters $\hat T_{sat},\hat \xi_{sat}$. 
 \end{itemize}
 \end{framed}

The strategy for Eve, in order to meet this level II criteria, will be to adjust not only the displacement $\Delta$, but also the gain $g$, in the saturation attack.

\subsection{Analysis and simulation results}

We will formalize two strategies for Eve and study numerically whether they can be used to meet the two criteria for the success of  the saturation attack, respectively.
We  use Eq.\eqref{eqt} and Eq.\eqref{eqex} to  perform numerical evaluation of $\hat T_{sat}$ and $\hat \xi_{sat}$, in order  to study the impact of the saturation attack.

 \subsubsection{Assumptions used in the numerical simulations}
\label{assumptions}
We have performed numerical simulations of the estimated excess noise $\hat \xi_{sat}$, the estimated channel transmission $\hat T_{sat}$, and the secret key rate under collective attack. We have chosen our simulation parameters in order to match typical parameters that can be achieved and chosen for the operation of an experimental CV-QKD system:
\begin{itemize}
\item Deployment over a dark fiber, with quantum channel wavelength in the C band and fiber attenuation coefficient $a=0.2dB/km$.
\item Total optical transmission (including homodyne detection finite efficiency) of Bob: $\eta=0.55$.

\item Linear detection limit of Bob's homodyne detection: $\alpha=20\sqrt{N_0}$.

\item Variance of the electronic noise $v_\mathrm{ele}=0.015N_0$, i.e. a result that can typically be achieved with a 10 MHz bandwidth homodyne detection and system clock rate of 1 MHz  \cite{Jouguet13}.
\item We have chosen a conservative value $\xi_{sys} =0.1$  for the system excess noise (equivalent excess noise at the input) in our simulations. This value is relatively high compared to some demonstrated experimental results in CV-QKD \cite{Jouguet13, KumarWDM2015} but it has been encountered in \cite{IR}, when performing the experimental demonstration of the intercept-resend attack . Adopting  a pessimistic value for system excess noise has in conservative and will not weaken our predictions concerning the power of the saturation attack on practical systems.
\item In a practical CV-QKD deployment, the value of Alice variance modulation $V_A$ depends on the link distance.  This is in particular due to finite reconciliation efficiency in practice. To achieve a high reconciliation efficiency in practical CV-QKD  ($\beta=0.95$), optimized error correction codes work at a fixed signal to noise ratio (SNR). Therefore Alice needs to optimize her modulation variance with respect to the distance in order to work at a given SNR.    We have followed the procedure described in Ref~\cite{Jouguet11} to choose Alice's variance with respect to distance in our numerical simulations.
\end{itemize}

 \subsubsection{Attack strategy I: Meeting level I criteria by varying $\Delta$}
 \label{stratone}

 Let us define strategy I more precisely:
 \begin{itemize}
  \item Eve implements the saturation attack as described in \ref{saturationattack}.
  \item Eve chooses a fixed gain value $G=g^2=2$  in order to compensate the loss due to heterodyne detection.
  \item By choosing the value of $\Delta$, Eve bias the excess noise estimation of Alice and Bob below the null key threshold,  so that Alice and Bob can obtain a positive key rate.
 \end{itemize}

  \begin{figure}[t]
        \centering
\subfigure[\, Estimated excess noise $\hat \xi_{sat}$ versus $\Delta$ for different distances.]{
\begin{minipage}[b]{0.45\textwidth}
\includegraphics[width=1\textwidth]{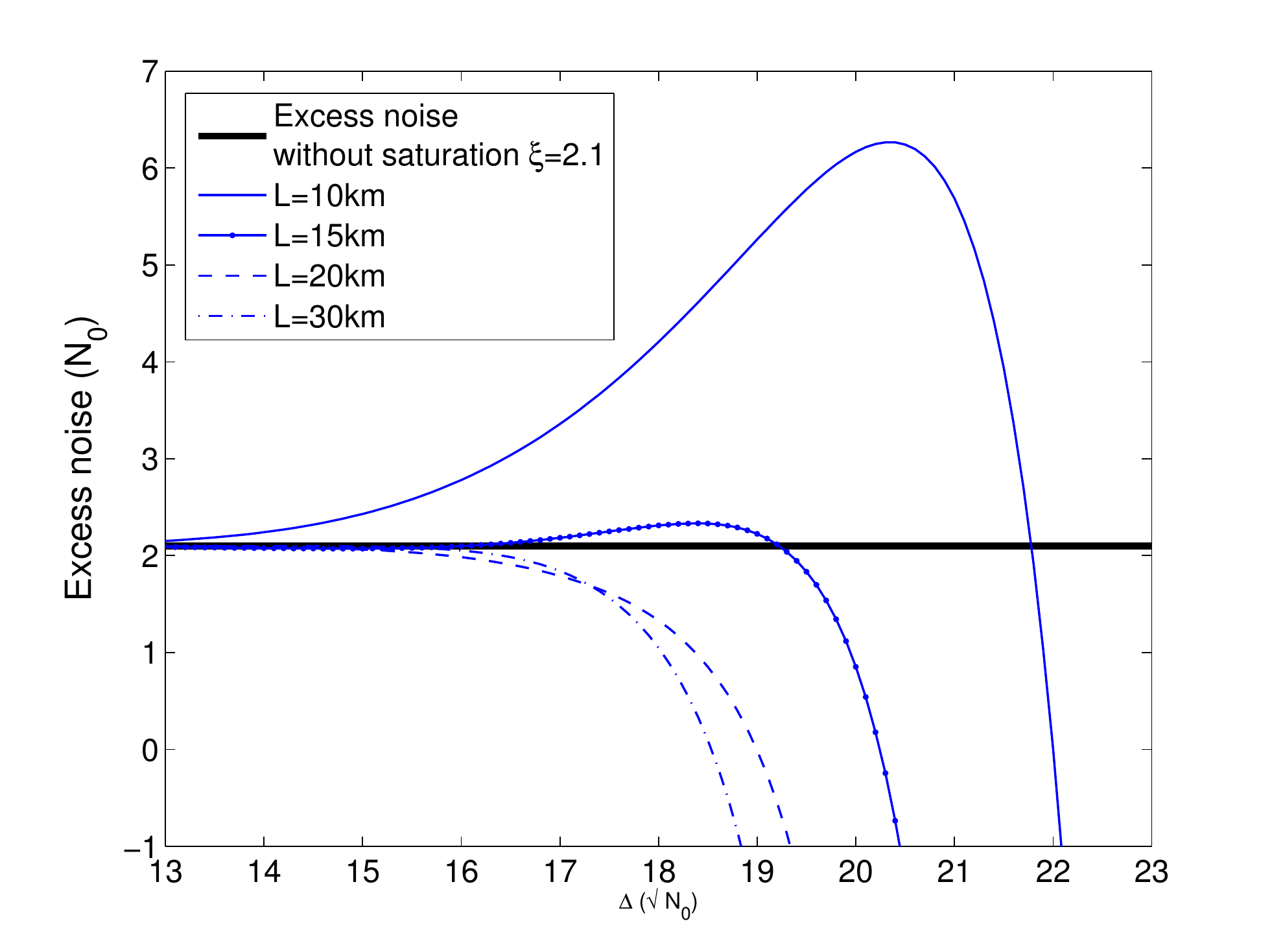}
\end{minipage}
}
\subfigure[\, Estimated transmission $\hat T_{sat}$ versus distance for different $\Delta$.]{
\begin{minipage}[b]{0.45\textwidth}
\includegraphics[width=1\textwidth]{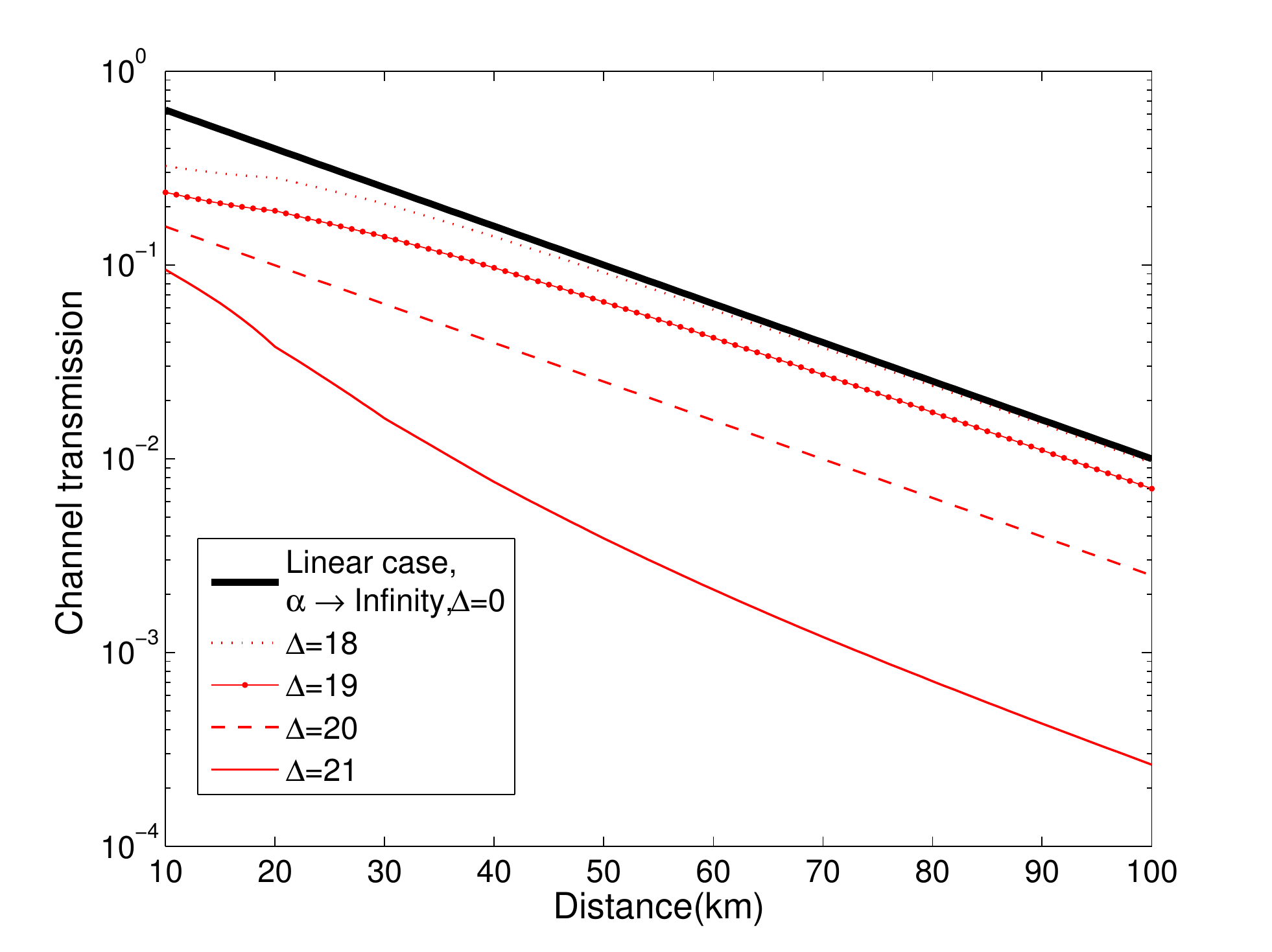}
\end{minipage}
}
        \caption{Simulations of estimated excess noise and channel transmission under attack strategy I. Simulation parameters: see \ref{assumptions}. }
        \label{T}
     \end{figure}

 The key idea of this strategy is that, for a given distance with the knowledge of $Var(X_{B_{lin}})$, Eve can manipulate  $\hat \xi_{sat}$ by  changing $\Delta$. More importantly, Eve needs to manipulate the excess noise evaluation so that   $\hat \xi_{sat}$ falls below the null key threshold, but remains positive, to meet the level I success criteria. $\hat \xi_{sat}$ is a function of $\Delta$ (Eq.\eqref{eqt}), the behavior of  $\hat \xi_{sat}$  versus  $\Delta$ is shown in Fig.\ref{T} (a). Under the linear model, the total estimated excess noise under a  full intercept-resend attack  is $\hat \xi_{lin}=\xi=2.1$ including 0.1 technical noise (black curves in Fig.\ref{T} (a)). With such an excess noise, no key rate can be established by Alice and Bob. However, $\hat \xi_{sat}$ can be manipulated by changing the value of $\Delta$.  In Fig.\ref{T} (a), for long distance (i.e. above 20 km) $\hat \xi_{sat}$ always decreases when   $\Delta$ increases.  Especially when $\Delta$ is close to $\alpha$,  $\hat \xi_{sat}$ is significantly reduced, which agrees with the analysis in section VII.A.   For short distance (i.e. below 15 km), when  $\Delta$ increases , $\hat \xi_{sat}$ first increases then decreases, but  $\hat \xi_{sat}$  can still become arbitrary small when $\Delta$ is large enough. Importantly, from Fig.\ref{T} (a), we can see that Eve can obtain an arbitrary small value of $\hat \xi_{sat}$ by manipulating $\Delta$ at any distance, which agrees with the analysis in section VII.A.2.
  
 A drawback of this strategy I of saturation attack is that the estimated channel transmission can be strongly reduced, i.e. we can have $\hat T_{sat} \ll T$ (Eq.\eqref{eqt}). In Fig.\ref{T} (b) we plot the estimated channel transmission in log scale versus distance, in which  the black curve  is the estimated channel transmission ($T$) versus distance in absence of attack while the other curves are estimated channel transmission ($\hat T_{sat}$) under the saturation attack.   We can see that the estimated channel transmission can be strongly reduced in comparison to the actual transmission  in absence of attack.  This is especially if $\Delta$ is large, which will be the case, as we can see on Fig.\ref{T} (a) for short links, where it is necessary to use large value of $\Delta$ to  effectively reduce the excess noise estimation and meet criteria I. Hence even though the attack strategy I can always been mounted, it may lead to use large displacement value $\Delta$ (typically or even beyond the saturation limit $\alpha$ set to 20 in our simulations). This will  strongly reduce the effective transmission of the channel $\hat T_{sat}$ and therefore the achievable secret key rate.

\subsubsection{Attack strategy II :  Meeting level II criteria by varying $\Delta$ and $g$}

   
   As we have just discussed, inducing the saturation of the homodyne detection (through the displacement $\Delta$) can lower the correlation between Alice and Bob's data, which will result in the decrease of the estimated  channel transmission $\hat T_{sat}$ (Fig.\ref{T}(b)) and thus also of the achievable secret key rate by performing the GMCS CV-QKD protocol over the the channel. 
   
 However, as already stated in  \ref{definecriteria},  there are many practical cases where Alice and Bob may monitor, or at least perform some kinds of consistency check on the estimated transmission and could therefore identify a problem if the estimated transmission, that becomes $\hat T_{sat}$ under the attack, is significantly smaller than the value of $T$ they expect. 

This motivates to define a second attack strategy, capable of meeting level II criteria: perform the saturation attack, obtain a positive key rate while leaving the estimation the channel transmission unchanged.

Level II criteria can clearly not been achieved with solely by varying the displacement $\Delta$ applied on the coherent states. However, the intercept-resend that is part of the saturation attack leaves an additional degree of freedom: Eve can apply rescale the value of the resent quadratures (classical result obtained after the heterodyne detection) by a gain $g$ that she can also freely choose.

 To meet the success criteria II, we will study a second strategy is similar to strategy I except for  the second step  where the gain $g$ will be set according to Eq.(\ref{eqg}), In which $Var(X_{B_{lin}})$ is given by Eq.\eqref{eqvbn}.

As a matter of fact if Eq.\eqref{eqg} is verified, then  $\langle X_AX_{B_{sat}}\rangle=\dfrac{1}{2}  \langle X_A^2\rangle t$ and we will thus have $\hat T_{sat}=T$,  which guarantees that the channel transmission estimation for Alice and Bob is not biased. 

\begin{equation}
\dfrac{2\sqrt{2}}{g}-1=\mathrm{erf}(\frac{\alpha-\Delta}{\sqrt{Var(X_{B_{lin}})}})
\label{eqg}
\end{equation}

\begin{figure}[t]
       
              \centering
      \subfigure[\, Gain $g$ verifying Eq. \ref{eqg} as a function of $\Delta$.]{
      \begin{minipage}[b]{0.45\textwidth}
      \includegraphics[width=1\textwidth]{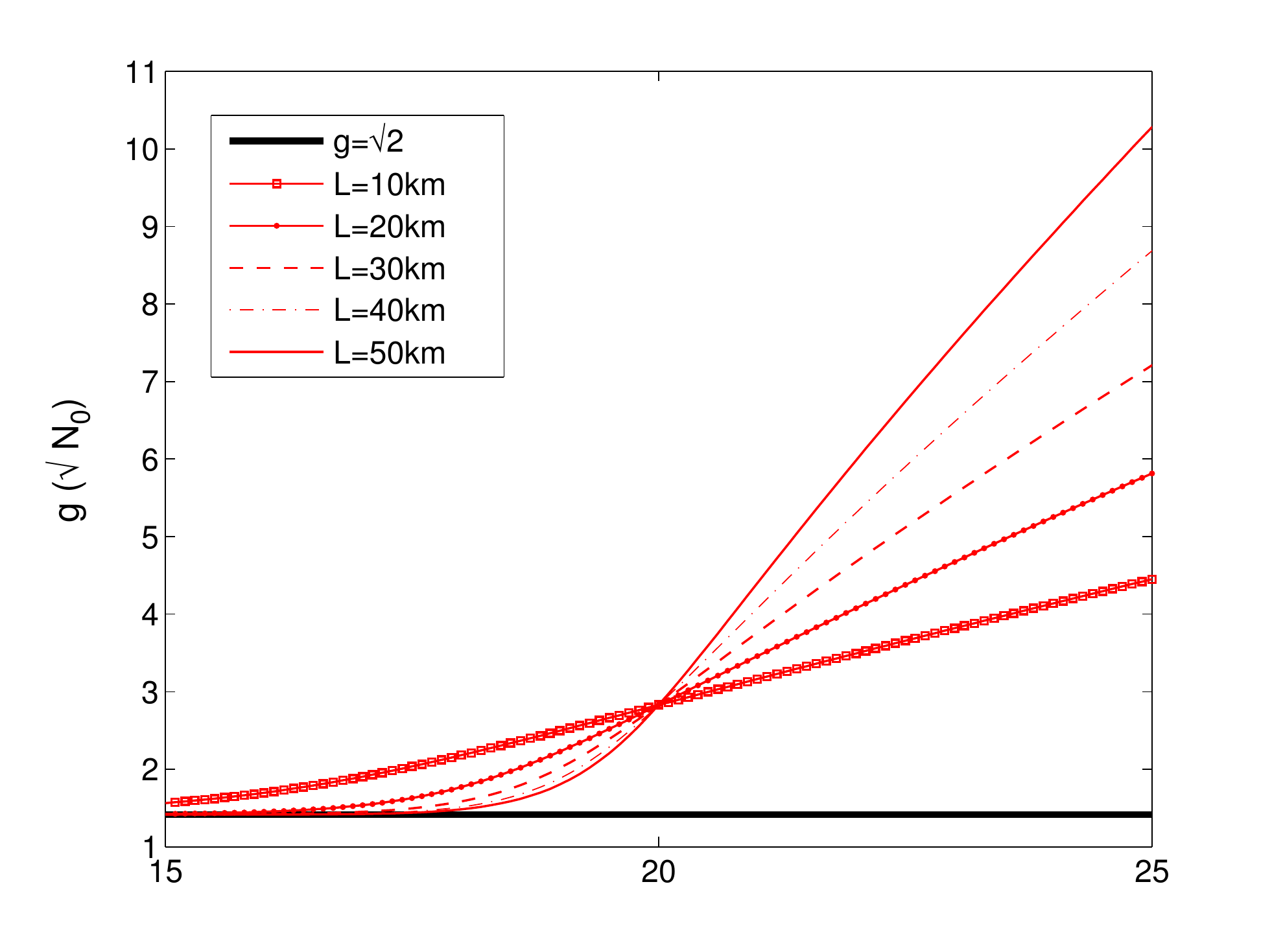}
      \end{minipage}
      }
      \subfigure[ \, Estimated excess noise as a function of $\Delta$ for different link lengths.]{
      \begin{minipage}[b]{0.45\textwidth}
      \includegraphics[width=1\textwidth]{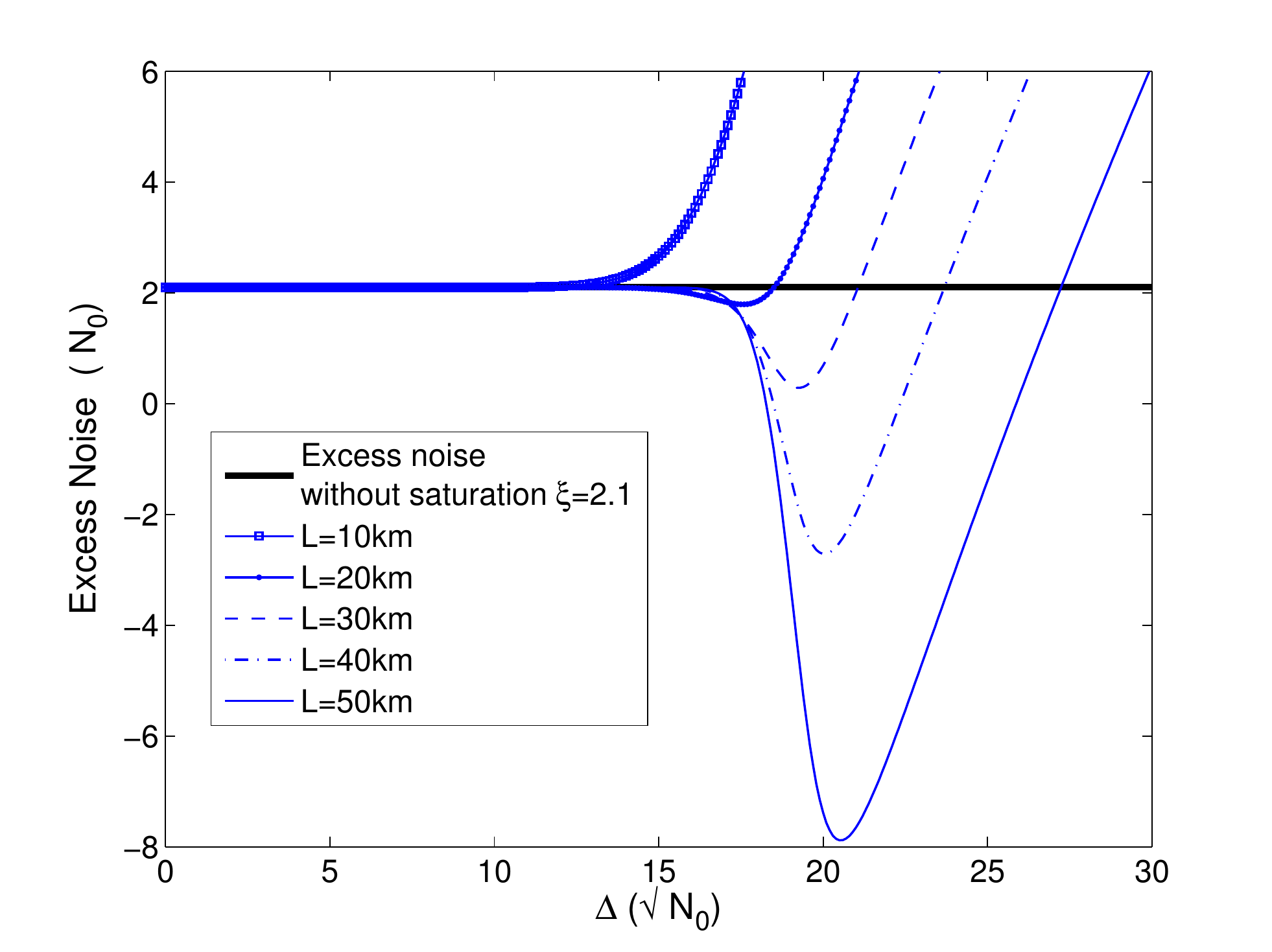}
      \end{minipage}
      }
              \caption{Simulations under attack strategy II Part 1. Simulation parameters: see \ref{assumptions}.}
              \label{ex}
           \end{figure}

  \begin{figure}[t]
 \centering
               \subfigure[\, Excess noise versus $\Delta$ and distance, solid lines with marks: estimated excess noise $\hat \xi_{sat}$ , dashed lines:  null key threshold $\xi_{null}$.]{
               \begin{minipage}[b]{0.45\textwidth}
               \includegraphics[width=1\textwidth]{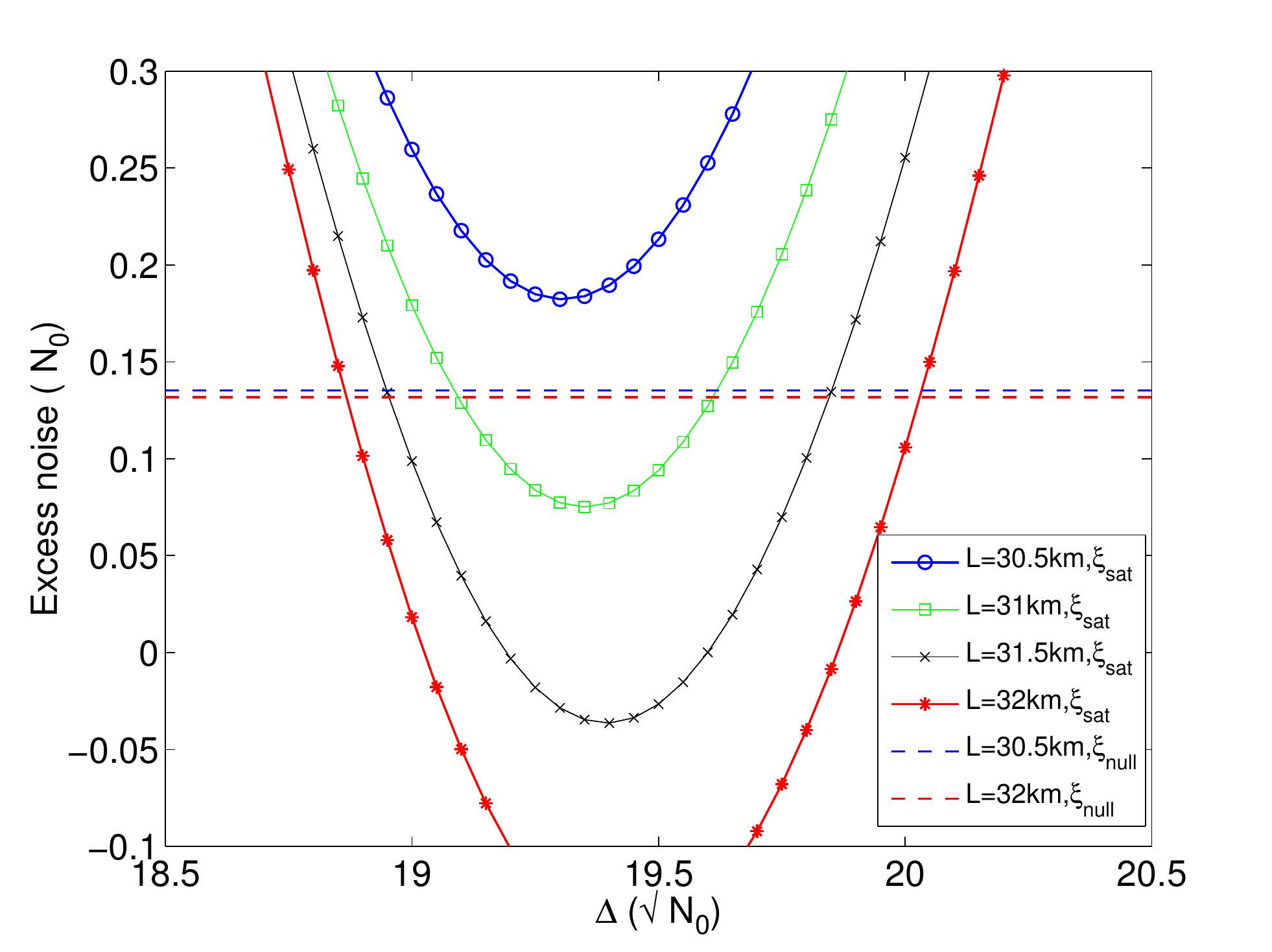}
               \end{minipage}
               }
               \subfigure[\, Key rate versus distance for different value of $\Delta$. No attack is possible for links shorter than 31 km under criteria II (see text).]{
               \begin{minipage}[b]{0.45\textwidth}
               \includegraphics[width=1\textwidth]{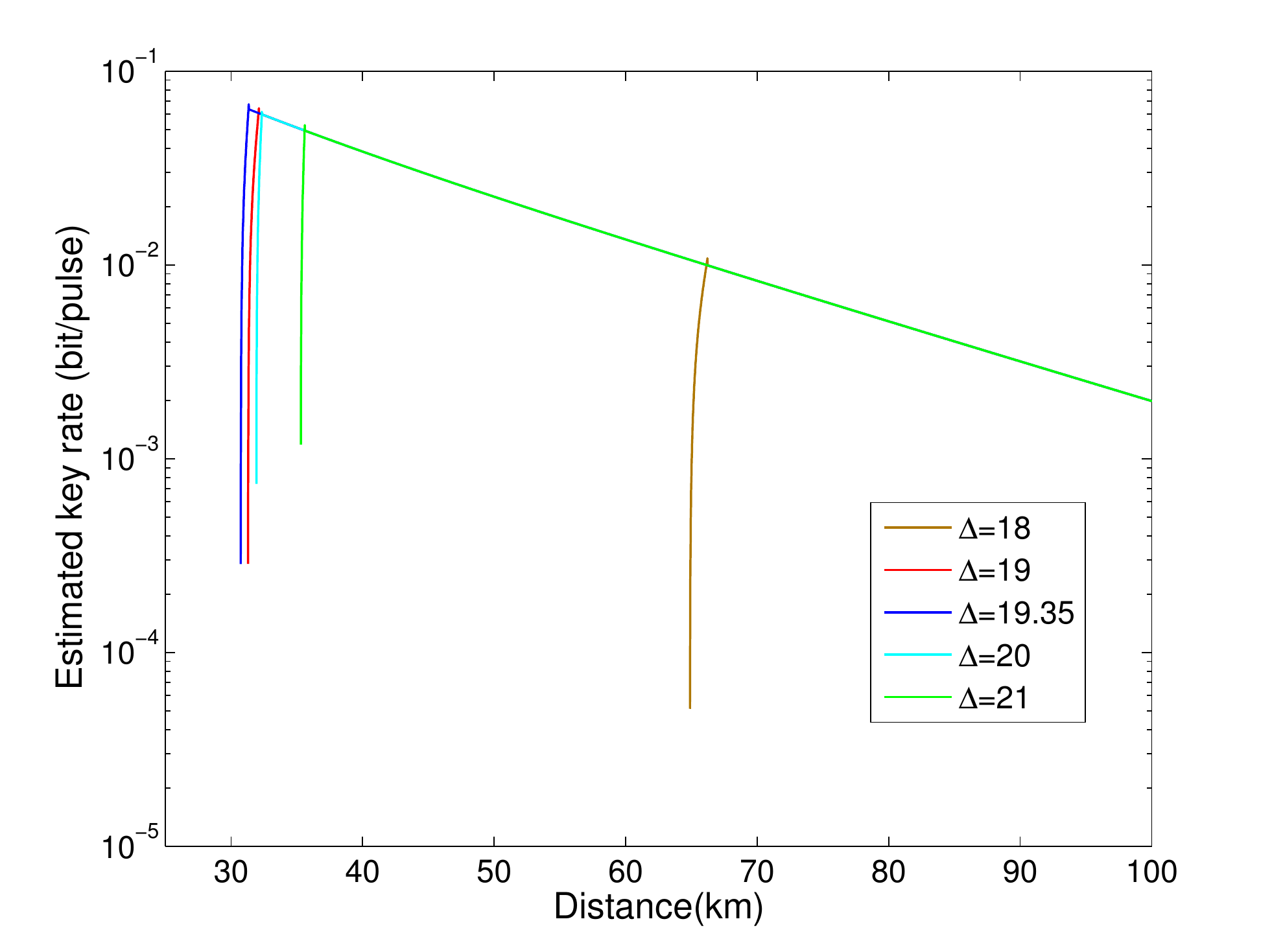}
               \end{minipage}}
\caption{Simulations under saturation attack strategy II Part 2. Simulation parameters: see \ref{assumptions}.} 

               \label{key2}
            \end{figure}


$g$ can now been considered as a function of $\Delta$, as displayed in Fig.\ref{ex}(a). Furthermore, in order to see whether we can meet criteria II and have a full security break with this new choice of $g$ we still need to analyze the estimated of  excess noise and secret key rate. By taking the $g$ solutions of Eq.\eqref{eqg} into account, the behavior of $\hat \xi_{sat}$ versus $\Delta$ is shown in Fig.\ref{ex} (b). 

We can see that if the distance is longer than 30 km it is always possible to reduce $\hat \xi_{sat}$ close to zero by choosing a value of $\Delta$ close to $\alpha$, and thus to have an attack meeting criteria II. On the other hand, if the distance is smaller than approximately 30 km, it will not be possible to meet the attack success criteria II and to jointly maintain the estimate of the channel transmission unchanged and have a positive key rate while launching the saturation attack. Thus the capacity to launch a successful saturation attack under success criteria II is dependent on the distance, as it can be seen  to Fig.\ref{ex} (b).

We also need to study the condition for  $\hat \xi_{sat}< \xi_{null}$ as we previously did in \ref{stratone}. However  the analysis is now simpler, since the estimated channel transmission is not biased, the null key threshold does not depend on the attack parameter $\Delta$ and only varies with distance. In  Fig.\ref{key2} (a) we enlarge the scale of Fig.\ref{ex} (b) and compare the estimated excess noise to  the null key threshold for different distances. As we can see, when the distance reaches 31 km,  the condition $\hat \xi_{sat}< \xi_{null}$ can only be satisfied with a choice of $\Delta \simeq 19.5$  and level II criteria conditions cannot be met for smaller distances.

We also estimate the secret key rate of Alice and Bob versus distance (Fig.\ref{key2} (b)). A set of parameters $\Delta$ and $g$ can always be found to meet success criteria II as long as the distance is large enough (larger than $31$km with our simulation parameters, detailed in \ref{assumptions}). Since the estimated channel transmission $T$ is unchanged, the estimated key rate will be identical to the key rate in absence of attack. Hence reaching strategy II, although it cannot be launched on short channels (high transmission) is a more powerful and more convincing strategy.

\section{Counter-measures against the saturation attack} \label{Counter measure}


 The vulnerability to the saturation attack, studied in previous sections, is related to the fact that the first moment (mean value) of the measured quadratures are by default not monitored in a CV-QKD protocol and can therefore be freely manipulated by an attacker, opening a practical security loophole. The essence of a counter-measure against the saturation attack will therefore consist in adding some test procedure to the CV-QKD protocol, in order to rule out the possibility that the detector saturates, i.e. that some input optical state has a quadrature $X_{in}$ larger than $\alpha$, characterizing the linear range of the detection. We present what could be this test procedures, also called counter-measures agains saturation. They range from post-selection tests, that can be implemented without any modification of CV-QKD hardware, to more structural modifications of the protocols, requiring extra hardware. Importantly we first recall that most counter-measure rely on some pre-existing calibration of the detector.



\paragraph{Pre-requisite: calibration and characterization of the homodyne detection linear range}



The scope of this article is restricted to Prepare and Measure (P$\&$M) CV-QKD, where the detector can be considered trusted, i.e. not influenced by the eavesdropper. In this context, Alice and Bob, the legitimate users, can rely on a (trusted) calibration of the detector, and in particular on a characterization of the detector linear range.
The homodyne detection is a phase-sensitive device that transforms an input optical state of quadrature $X_{in}$, measured with respect to the phase reference (local oscillator) into a measured voltage $V_{out}$. For an unsaturated homodyne detection, the relation between $X_{in}$ and $V_{out}$ is linear, with a linear gain that depends on several parameters such as the local oscillator amplitude, the optical loss, the mode matching and the electronic gain of the electronics (a transimpedance circuit is commonly used to perform low-noise measurement of the small differential photocurrent associated with $X_{in}$). All these parameters are not easy to measure independently, but we can calibrate them globally by measuring (offline, for example before launching the CV-QKD protocol) the variance of the homodyne output voltage $V_{out}$ when the input is vacuum, possibly for different values of the local oscillator power. This corresponds to measuring the shot noise variance (in voltage) $N_{0,V}$. Quadrature measurements $X_{out}$ are then usually expressed in square-root of shot noise units, which means that they are renormalized, based on shot noise calibration: $X_{out} \equiv V_{out}/\sqrt{N_{0,V}}$.\\
$X_{out}$, expressed in square-root of shot noise units, corresponds to the quadrature measurement we want to perform with our homodyne detection. If the input optical state is also expressed in square-root of shot noise units then wen have, after calibration, $X_{in} = X_{out}$, {\it provided the detection is linear}.

However, as illustrated in Fig. \ref{mean}, a realistic detector saturates and in practice the linearity between $X_{in}$ and $X_{out}$ can only be guaranteed over a finite range. 
We have called $\alpha$ the linearity bound such that $\forall |X_{in}| < \alpha \; \; \; X_{out} = X_{in}$. $\alpha$ is a characteristics of the homodyne detector, for a given local oscillator power. In practice, the local oscillator power should be chosen not too large such that the saturation limit is much larger than the shot noise, i.e. if $\alpha$ is expressed in shot noise units,  we want to have $\alpha^2 \gg 1$. Inversely, the local oscillator power should be chosen not too small, so that the variance of electronic noise, in shot noise unit, is much smaller than the shot noise variance: $v_{elec} \ll 1$

\paragraph{From an intuitive but faulty counter-measure to an efficient counter-measure based on ``radical post-selection''}
A simple (but faulty) counter-measure would consist in post-selecting quadrature measurement results provided they fall in a ``confidence interval'' where the homodyne detection is known to be linear, i.e. typically if they fall within an interval of the form $[- (\alpha-\beta), \alpha-\beta ]$. $\beta$ can be seen as a confidence margin, with $0 < \beta < \alpha$. 
We can however see that such counter-measure would trigger new problems: it would give Eve the possibility to influence which data is post-selected and which one is not, just by controlling the displacement value. The post-selected data would not be Gaussian and no security could be guaranteed. This observation has motivated the development of the counter-measure based on Gaussian post-selection that we will detail at the end of this section. 

A more radical counter-measure is however possible, that consists in discarding measurement data blocks if  {\it any} of the measured data falls out of the confidence interval $[- (\alpha-\beta), \alpha-\beta ]$.  Quadrature measurements and parameter estimation are in practice realized over blocks of large size in order to limit finite-size effect (for example $10^8$ in \cite{Jouguet13}). If the detector has been properly calibrated and if its characteristics are stable over time, then counter-measure based on what we will call ``radical post-selection'' will guarantee by construction that provided it passes the test, a data block has been acquired with a detector operated in a linear regime. The drawback of this radical post-selection procedure is that it might lead to discard some or possibly almost all data blocks if the confidence interval is not large enough (and/or not centered) with respect to the impinging optical quadrature variance. This counter-measure will efficiently protect against saturation attacks and is easily implementable. Provided the distribution of the input quadrature $X_{in}$ is centered, the confidence interval should typically be larger than six standard deviations, i.e. $\alpha-\beta > 6 \, \sqrt{Var(X_{in})}$, so that the probability (per measurement event) to have a saturation is below $2 \, 10^{-9}$ and hence the probability to discard ``good'' data blocks remain relatively small. On the other hand, if the linearity domain of the homodyne detection is not large enough, then this countermeasure might strongly affect the effective key rate ,  even in the absence of any attack, which has motivated us to propose a more refined counter-measure.

\paragraph{Gaussian post-selection}

We now propose a refined version of the ``radical-post-selection''. This new method retains the important advantage of being implementable ``at the software level'' by modifying the post-processing stage. It can moreover cope with detector saturation in a more gentle way than throwing away the entire data block, as soon as a saturated measurement is detected, as it is the case with radical post-selection.

The method is based on performing a Gaussian post-selection of the measurement data. The key idea of such method is to extract a set of (almost) Gaussian distributed data among the raw measurement data, and to adjust the parameters of the post-selection so that post-selected data falls almost certainly within the (calibrated) linearity domain of the homodyne detector. Calling $g(x)$ the probability distribution of data points after post-selection, the goal of the post-selection procedure is to choose the parameters of the non-normalized Gaussian filter $g(x)$ (mean value $\mu_g$, variance $\sigma_{g}^2$) under the following constraints :
\begin{enumerate}
\item $g$ is a post-selection function which implies there are less points after than before post-selection: $\forall x \in [-\alpha, \alpha],  0 \leq g(x) \leq f(x)$,
\item post-selected data should be almost Gaussian, i.e. the support of $g(x)$ should be almost contained in the linearity domain:  $\int_{-\alpha}^{\alpha}g(x) dx \simeq  \int_{-\infty}^{\infty} g(x) dx $ 
\item the number of post-seletect points, $ N^{'} \equiv \int_{-\alpha}^{\alpha} g(x) dx  $, should be maximized, under the two previous constraints.
\end{enumerate}

Performing the Gaussian post-selection first consists in binning the measured data (size $N$). Calling $f(x)$ the normalized distribution distribution of the raw data (quadrature measurement data) and considering bins centered on measurement result $x$ and of width $ \delta x$, there are approximately $N f(x) \delta x $ raw data points falling within a bin. The Gaussian post-selection consists in randomly selecting a fraction $g(x)/f(x)$ of those points  falling within the bin, and throwing away the others. After this procedure, applied to each bin, the post-selected data will have a probability distribution close to $g(x)$ (provided we have large enough data blocks and use small enough bins, so that finite size effects remain small).

\begin{figure}[t]
    
           \centering
     {\includegraphics[width=0.5\textwidth]{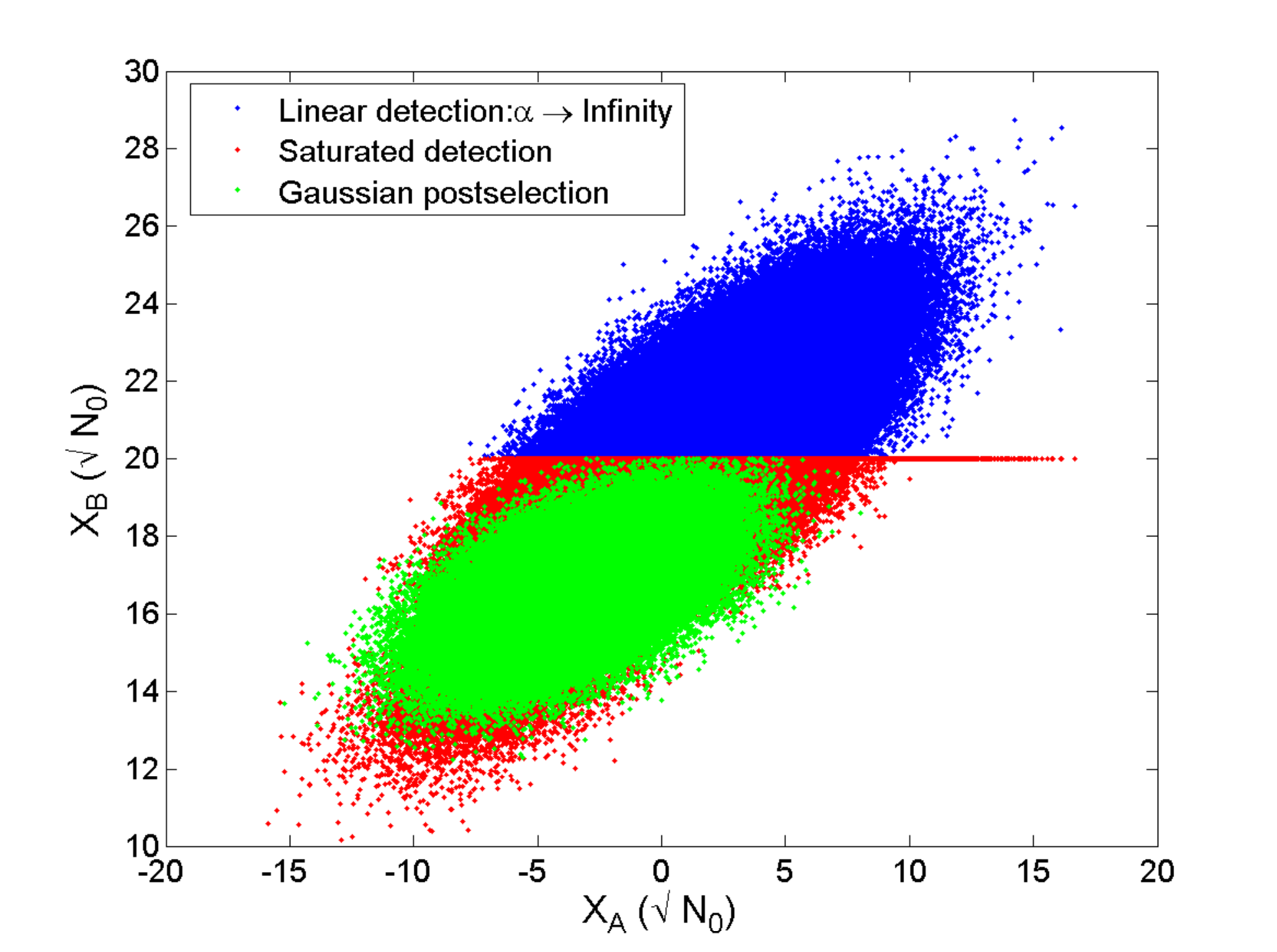}}
           \caption{Simulation of  Gaussian post-selection. Blue dots: simulated experimental data with infinite linear detection limit, $\Delta=19.2\sqrt{N_0}$ $L=20km$, $V_A=11.58N_0$, data number $N=10^6$;     Red dots:  simulated experimental data with  linear detection limit $\alpha=20 \sqrt{N_0}$, other parameters are same as blue ones.  Green dots: Gaussian post selected data among the red dots;   Gaussian post-selection parameter, $\sigma_{g}^2 = 2.5N_0$, $\mu_{g}=16.55N_0$; number of post-selected points : $N^{'}=15.37\%N$. Other simulation parameters refer to section.\ref{assumptions}.}
           \label{gs}
        \end{figure}

In order to illustrate how this Gaussian post-selection method could be realized in practice, we have simulated a CV-QKD experiment affected by saturation. The results are displayed on Fig.\ref{gs}.  The total number of points is $N=10^6$.  We have assumed a lossy channel between A and B, a displacement $\Delta$ at B, and a homodyne detection affected by saturation, as described in \ref{model}. We have used the following numerical values: channel distance 25 km, Alice variance $V_A=11.58N_0$, displacement $\Delta=19.2\sqrt{N_0}$. 
The blue dots would correspond to measurement results with a perfect homodyne detection (no saturation) while the red dots correspond to the results for the realistic detection, affected by saturation, with a linearity limit characterized by $\alpha=20 \sqrt{N_0}$.  The green dots correspond to the Gaussian post-selected data, applying the procedure detailed above. In order to find the parameter of the Gaussian filter $g(x)$ we have first removed the data points falling outside of the linearity domain $[ -\alpha, \alpha]$ and then built the histogram of $X_{B_{i}}$ with bin size $\delta x =0.1\sqrt{N_0}$. We could then estimate de probability distribution $f(x)$. The essential remaining step was to choose the parameters (variance, mean value, and amplitude $A$ ) of the Gaussian filter 
\begin{equation}
g(x) \equiv \dfrac{A}{\sqrt {2\pi} \sigma_{g} } exp({-\frac{(x-\mu_g)^2}{2 \sigma_{g}^2  }})
\end{equation}

We have optimized numerically these parameters to maximize the number of post-selected points $N^{'}$, under the constraints 1 and 2.\\ We have obtained a total number of post-selected points $N^{'}=15.37\%N$ (Green dots in Fig.\ref{gs}), while guaranteeing that the L2 distance between the normalized post selected data distribution and a perfect Gaussian distribution  is below $10^{-3}$. 


The Gaussian post-selection is more complex to implement than the radical post-selection, but has the advantage of allowing to generate some key, even in presence of moderate saturation. The post-selection also guarantees that post-selected points fall within the linearity domain of the detector (therefore guaranteeing linearity of the detector on these post-selected data points). Moreover the Gaussian post-selection also guarantees that the distribution of the input data, after post-selection, is Gaussian and thus that the post-selected data still implements the GMCS protocol, although with different channel parameters $T'$ and $\xi'$. As a consequence, provided $T'$ and $\xi'$ are compatible with secure key generation, some key can be distilled from the post-selected data, free from any security threat and attacks exploiting detector saturation.

\paragraph{Counter-measures relying on existing techniques, necessitating some additional hardware}

The two (radical and Gaussian) post-selection measures discussed above can be implemented without any additional hardware, which constitutes an important practical advantage. For completeness we also discuss here other ways to counter attacks related to saturation, that however all involve some changes not only on the protocol, but also on the hardware side.

As proposed and implemented in \cite{RobustShotNoiseM}, varying randomly the attenuation, on the signal port, at the input of Bob, allows to test the linearity of the homodyne output with respect to the signal input. The saturation attack exploits in particular the fact that the shot noise variance  is calibrated in absence of signal (attenuation $\eta =0$) while the total noise variance is measured  with no attenuation ($\eta =1$) and with saturation (when the attack is launched) As a consequence the excess noise is underestimated. Such attack cannot however be performed  if the total noise variance is estimated with more than 2 attenuation values and in particular is the linearity of the total noise with the attenuation is checked. This is precisely the countermeasure proposed in \cite{RobustShotNoiseM}, that constitutes an effective parade against the saturation attack. The drawback of this countermeasure is however that it requires another hardware element, bringing additional complexity, and also that it leads to attenuate the signal and thus reduce the key rate.

MDI CV-QKD can be used to perform QKD with untrusted detectors. It could therefore in particular be used to perform CV-QKD securely with practical homodyne detectors, subject of saturation. This would however  be at the expense of a significant increase of the experimental complexity (in particular phase locking of two distant lasers) and also at the expense of performance since only low to moderate losses can be tolerated in MDI CV-QKD \cite{MDICVQKD}.

Finally, one could notice that the saturation attack requires to strongly displace the mean value of the quadrature signal. The signals impinging on the homodyne detection must therefore have a high energy. The method proposed in \cite{FurrerPRA}, in another context, could then be used as a counter-measure: it consists in upper bounding, with an auxiliary homodyne detector (sometimes called watch-dog in other contexts \cite{watchdog}) the energy of the impinging signals. One limitation of this method is however that it is in general not possible to predict in which mode an attacker will try to send energy in order to saturate the detector. It can thus be very difficult to design in practice a watch-dog capable of detecting any attempts to saturate the detector.

\section{Conclusion}\label{conclusion}

We have studied quantitatively how the saturation attack can be used to compromise the security of practical CV-QKD systems. The main finding of our study is that the excess noise can be actively reduced by displacing the quadratures mean value of the coherent states received by Bob, and that this effect can compromise the security of Gaussian-modulated coherent state CV-QKD protocol, operated with practical detectors whose linearity response can only be guaranteed over a restricted domain of quadrature values.

We have proposed an explicit attack, called saturation attack, that combines displacement with a full intercept-resend attack. We have performed numerical simulations that show the feasibility of our attack under realistic experimental conditions. The saturation attack consists in strongly displacing the quadrature mean values, to induce saturation of Bob detector. Our attack is achievable with current technology and may impact the security all CV-QKD implementation, since any practical detectors is subject to saturation. An experimental demonstration of this attack is the topic of an ongoing work. 

 While all previous attacks on CV-QKD had focused on local oscillator manipulation and biasing excess noise evaluation, our attack has no influence on the local oscillator, and can thus not be ruled out by generating the local oscillator locally \cite{BQi2015, Soh2015}. It is therefore important to propose practical solutions against this attack, and we have presented in detail two effective counter-measures based on post-selection, that can be implemented without requiring any modification at the hardware level. This work illustrates the importance of putting under great scrutiny the hypothesis under which the security proof can be derived, but also illustrates that secure and yet still practical QKD implementations are within reach.


\section{Acknowledgments} 
R.A. thanks Fr\'ed\'eric Grosshans for stimulating discussions about counter-measures. This research is supported by the French National Research Agency, through the FREQUENCY (ANR-09-BLAN-0410) and by the European Union through the project Q-CERT (FP7-PEOPLE-2009-IAPP) and the ANR project  Quantum-WDM (ANR-12-EMMA-0034). H.Q. acknowledges support from the Institut Mines-T\'{e}l\'{e}com, Paris, France.



\clearpage
\appendix
\section{Calculation of the correlation under the saturation attack}
In order to clearly show the calculation, we consider $y_{sat}$, $y$, $x$ and $z$ as the notations of  $X_{B_{sat}}$, $X_{B_{lin}}$, $X_A$ and $X_N$ respectively. We use  $X_{B_{sat}}$ (Eq.\eqref{eqsatd}) to calculate the correlation $ Cov(X_A,X_{B_{sat}})$ under the saturation attack. We assume here $\alpha >> 1$  and consider $\Delta\geq 0$, while the analysis of $\Delta\leq 0$ is similar. The saturation model can be considered as:
\begin{equation}
    \begin{split}
     &y_{sat}=\alpha,  \qquad \qquad t\dfrac{g}{\sqrt 2}x+z+\Delta \geqslant \alpha \\
     &y_{sat}=t\dfrac{g}{\sqrt 2}x+z+\Delta,   \mid t\dfrac{g}{\sqrt 2}x+z+\Delta \mid<\alpha(\alpha>> 1,\Delta\geq 0)\\
     &y_{sat}=-\alpha, \qquad \qquad      t\dfrac{g}{\sqrt 2}x+z+\Delta \leqslant -\alpha
     \end{split}
     \label{eqeve}
\end{equation}
$x\sim \mathcal N (0,\sigma_x^2)$ and $z\sim \mathcal N (0,\sigma_z^2)$ are both centered  Gaussian variables with  probability density function $p_X(x)$ and $p_Z(z)$, respectively:
\begin{equation}
p_X(x)=\dfrac{e^{-\frac{x^2}{2\sigma_x^2}}}{\sqrt {2\pi}\sigma_x},\qquad  
p_Z(z)=\dfrac{e^{-\frac{z^2}{2\sigma_z^2}}}{\sqrt {2\pi}\sigma_z}.
\end{equation}
In which $\sigma_x^2=Var(X_A)$ and $\sigma_z^2=N_0 + \eta T \xi + v_\mathrm{ele}$.
By knowing $p_X(x)$ and $p_Z(z)$, we can calculate $Cov(x,y_{sat})$ with double integral of $x$ and $z$ in the domain $D_{xz}$. $D_{xz}$ is defined in Eq.\eqref{eqeve}: $-\alpha<\frac{tg}{\sqrt 2}x+z+\Delta<\alpha$, $\frac{tg}{\sqrt 2}x+z+\Delta\leq-\alpha$ and $\frac{tg}{\sqrt 2}x+z+\Delta\geq\alpha$.  A long but straight forward calculation of  $Cov(x,y_{sat})$ is presented as follows:
\begin{equation}
\begin{split}
& Cov(X_A,X_{B_{sat}}) =\langle xy_{sat}\rangle-\langle x\rangle\langle y_{sat}\rangle=\langle xy_{sat}\rangle\\&=\iint\limits_{D_{xz}}xyp_X(x)p_Z(z)dxdz\\&= \iint\limits_{-\alpha<\frac{tg}{\sqrt 2}x+z+\Delta<\alpha}(\frac{tg}{\sqrt 2}x^2+x\Delta +xz) p_X(x)p_Z(z)dxdz\\&
 +\iint\limits_{\frac{tg}{\sqrt 2}x+z+\Delta\leq-\alpha}-\alpha x p_X(x)p_Z(z)dxdz \\&+ \iint\limits_{\frac{tg}{\sqrt 2}x+z+\Delta\geq\alpha} \alpha x p_X(x)p_Z(z)dxdz.
  \end{split}
 \end{equation}

\begin{equation}
 \label{eq:covsat}
     \begin{split}
& Cov(X_A,X_{B_{sat}})=\\=&\dfrac{1}{2\pi \sigma_x \sigma_z }\int\limits_{-\infty}^{\infty}(\frac{tg}{\sqrt 2}x^2+x\Delta)e^{-\frac{x^2}{\sigma_x^2}}dx\int\limits_{-\alpha-\Delta-\frac{tg}{\sqrt 2}x}^{\alpha-\Delta-\frac{tg}{\sqrt 2}x}e^{-\frac{z^2}{2\sigma_z^2}}dz\\=&\dfrac{1}{2\pi \sigma_x \sigma_z }\int\limits_{-\infty}^{\infty}(\frac{tg}{\sqrt 2}x^2+x\Delta)e^{-\frac{x^2}{\sigma_x^2}}\sqrt{\dfrac{\pi}{2}}\sigma_z[\mathrm{erf}(\dfrac{\alpha+\Delta+\frac{tg}{\sqrt 2}x}{\sqrt {2}\sigma_z })\\&+\mathrm{erf}(\dfrac{\alpha-\Delta-\frac{tg}{\sqrt 2}x}{\sqrt {2}\sigma_z })]dx\\=&\dfrac{1}{2\pi\sigma_x}\sqrt{\dfrac{\pi}{2}}[\dfrac{tg}{\sqrt{2}}\int\limits_{-\infty}^{\infty}x^2e^{-\frac{x^2}{2\sigma_x^2}}dx\\&+\Delta\int\limits_{-\infty}^{\infty} \mathrm{erf}(\dfrac{\alpha-\Delta-\frac{tg}{\sqrt 2}x}{\sqrt {2}\sigma_z })xe^{-\frac{x^2}{2\sigma_x}}dx]\\=&\dfrac{tg}{2\sqrt 2\pi \sigma_x}\sqrt{\dfrac{\pi}{2}}\sqrt{2\pi}\sigma_x^3 \\&+\dfrac{tg}{2\sqrt 2\pi \sigma_x}\sqrt{\dfrac{\pi}{2}}\sqrt{2\pi}\sigma_x^3 \mathrm{erf}(\dfrac{\alpha-\Delta}{\sqrt{2(\sigma_z^2+\frac{t^2g^2}{2}  \sigma_x^2)}})\\=&   \dfrac{tg}{2\sqrt 2}\sigma_x^2[1+\mathrm{erf}(\dfrac{\alpha-\Delta}{\sqrt{2(\sigma_z^2+\frac{t^2g^2}{2}  \sigma_x^2)}})]
\end{split}
\end{equation}
Thus we can conclude that
\begin{equation}
\begin{split}
Cov(X_A,X_{B_{sat}})= \dfrac{tg}{2\sqrt 2}\langle X_A^2\rangle[1+\mathrm{erf}(\dfrac{\alpha-\Delta}{\sqrt{2 Var(X_{B_{lin}})}})].
 \end{split}
\end{equation}
In which,   the error function $\mathrm{erf}(x)$ is defined as:
\begin{equation}
\mathrm{erf}(x)=\frac{2}{\sqrt{\pi}}\int_0^x e^{-t^2}\,dt.
\end{equation}
And we have used the integral formulas of $\mathrm{erf}(x)$ provided in \cite{Ng1969}. In Eq.\eqref{eq:covsat}, $Var(X_{B_{lin}})=\sigma_z^2+\frac{t^2g^2}{2}  \sigma_x^2$ is variance of Bob with no saturation. In this calculation, the integrals of the odd functions with symmetric bounds $(-\infty,\infty)$ are equal to zero.

\section{Calculation of the variance of Bob under the saturation attack}
In order to calculate the variance of Bob under the saturation attack, we use the step function $\theta(x)$  which is defined as:
    \begin{equation}
    \label{theta}
    \theta(x)=\begin{cases}
    1, x\in [0,\infty)\\
    0, x\in (-\infty,0]
    \end{cases}
    \end{equation}
With  Eq.\eqref{theta} we can transform  Eq.\eqref{eqeve} into:
\begin{equation}
    \begin{split}
    y_{sat}&=y\theta(y+\Delta+\alpha)\theta(-y-\Delta+\alpha)\\&+\alpha[1-\theta(y+\Delta+\alpha)\theta(-y-\Delta+\alpha)]\\&\approx\alpha+(y+\Delta-\alpha)\theta(-y-\Delta+\alpha)\\&=\alpha+(y-\varepsilon)\theta(-y+\varepsilon).
    \end{split}
\end{equation}
in which:
\begin{align}
 \varepsilon&=\alpha-\Delta (\alpha>0,\Delta\geq 0),\\
y&=t\dfrac{g}{\sqrt 2}x+z.
 \end{align}
Since $x$ and $z$ are both Gaussian variables, $y$ is also a Gaussian variable ($y\sim \mathcal N (0,\sigma_y^2)$), with its probability function  $p_Y(y)=\frac{e^{-\frac{y^2}{2\sigma_y^2}}}{\sqrt {2\pi}\sigma_y}$ and $\sigma_y^2 = Var(X_{B_{lin}})$ is the variance of Bob under linear detection. In order to estimate $ Var(X_{B_{sat}})= Var(y_{sat})=\langle y_{sat}^2\rangle-\langle y_{sat}\rangle^2$, we need to calculate $\langle y_{sat}\rangle$ and 
$\langle y_{sat}^2\rangle$, respectively:
\begin{align}
\langle y_{sat}\rangle&=\alpha+\langle(y-\varepsilon)\theta(-y+\varepsilon)\rangle=\alpha+C,\\ 
\langle y_{sat}^2\rangle&=\langle\alpha^2+2\alpha(y-\varepsilon)\theta(-y+\varepsilon)+(y-\varepsilon)^2\theta(-y+\varepsilon)\rangle\\&=\alpha^2-2\alpha C+D.
\end{align}
In which $C$ and $D$ are equal to $\langle(y-\varepsilon)\theta(-y+\varepsilon)\rangle$ and $\langle(y-\varepsilon)^2\theta(-y+\varepsilon)\rangle$, and can be calculated as follows:
\begin{align}
C&=\int\limits_{-\infty}^{\infty}p_Y(y)(y-\varepsilon)\theta(-y+\varepsilon)dy\\
&=\int\limits_{-\infty}^{\infty}p_Y(y'+\varepsilon)y'\theta(-y')dy'=\int\limits_{-\infty}^{0}p_Y(y'+\varepsilon)y'dy'\\&=-[\dfrac{\sigma_y}{\sqrt{2\pi}}e^{-\frac{\varepsilon^2}{2\sigma_y^2}}+\dfrac{\varepsilon}{2}+\dfrac{\varepsilon}{2}\mathrm{erf}(\dfrac{\varepsilon}{\sqrt{2}\sigma_y})]
\end{align}

\begin{align}
D&=\langle(y-\varepsilon)^2\theta(-y+\varepsilon)\rangle=\int\limits_{-\infty}^{\infty}p_Y(y)(y-\varepsilon)^2\theta(-y+\varepsilon)dy\\&=\int\limits_{-\infty}^{\infty}p_Y(y'+\varepsilon)y'^2\theta(-y')dy'=\int\limits_{-\infty}^{0}p_Y(y'+\varepsilon)y'^2dy'\\&=\dfrac{\varepsilon\sigma_y}{\sqrt{2\pi}}e^{-\frac{\varepsilon^2}{2\sigma_y^2}}+\dfrac{\varepsilon^2+\sigma_y^2}{2}[1+\mathrm{erf}(\dfrac{\varepsilon}{\sqrt{2}\sigma_y})].
\end{align}
We have used  $y'=y-\varepsilon$ in the calculations of  $C$ and $D$. Provided with $C$ and $D$, we can calculate $Var(y_{sat})$:
\begin{align}
\label{eq:vbsat}
\begin{split}
Var(y_{sat})&=\langle y_{sat}^2\rangle-\langle y_{sat}\rangle^2\\
&=\alpha^2-2\alpha C+D-(\alpha+C)^2=D-C^2\\&=\sigma_y^2[\dfrac{1+\mathrm{erf}(\frac{\varepsilon}{\sqrt{2}\sigma_y})}{2}-\dfrac{e^{-\frac{\varepsilon^2}{\sigma_y^2}}}{{2\pi}}]-\dfrac{\varepsilon\sigma_y}{\sqrt{2\pi}}\mathrm{erf}(\frac{\varepsilon}{\sqrt{2}\sigma_y})e^{-\frac{\varepsilon^2}{2\sigma_y^2}}\\&+\dfrac{\varepsilon^2}{4}[1-\mathrm{erf}^2(\frac{\varepsilon}{\sqrt{2}\sigma_y})]\\&=  Var(X_{B_{lin}})(\dfrac{1+A}{2}-\dfrac{B ^2}{2\pi})\\&-(\alpha-\Delta)\sqrt{\dfrac{ Var(X_{B_{lin}})}{2 \pi}}A*B+\dfrac{(\alpha-\Delta)^2}{4}(1-A^2)
\end{split}
\end{align}
in which:
\begin{align}
A = \mathrm{erf}(\dfrac{\alpha-\Delta}{\sqrt{2 Var(X_{B_{lin}})}}), B=e^{-\frac{(\alpha-\Delta)^2}{2 Var(X_{B_{lin}})}}
\end{align}


\end{document}